\documentclass[a4paper, 8pt, notitlepage]{article}

\usepackage[top=25mm,bottom=25mm,left=25mm,right=25mm]{geometry}
\usepackage{multirow}
\usepackage{dsfont}
\usepackage{multicol}
\usepackage{amssymb,amsmath}
\usepackage{graphicx}
\usepackage{url}
\usepackage{setspace} 
\usepackage{titlesec}
\usepackage[utf8]{inputenc}
\usepackage{wrapfig}
\usepackage{xcolor}
\usepackage{mhchem}
\usepackage{caption}
\usepackage[numbers]{natbib}
\usepackage{hyperref}
\usepackage{cleveref}
\crefname{figure}{Fig.}{Figs.}
\Crefname{figure}{Fig.}{Figs.}

\title{Near-Field Characterisation of Guided Modes in WS\(_2\) Nanobeams and Quasi-Bulk Crystals}
\author{Zara S. Taylor}
\begin{document}
\begingroup  
  \centering
  \LARGE Near-Field Characterisation of Guided Modes in WS\(_2\) Nanobeams and Quasi-Bulk Crystals
  \vspace{2mm}
  \normalsize \\ Zara S. Taylor\(^{1, *}\), Luke M. Hallacy\(^1\), Xuerong Hu\(^1\), Oliver T. Williams\(^1\), Simone Strohmair\(^2\), Fabian Felixberger\(^2\), Alexander J. Knight\(^1\), Timothy Chester-Parsons\(^1\), Luke R. Wilson\(^1\), Alexander I. Tartakovskii\(^{1, **}\) \\ 
  \vspace{-1mm}
    \(^1\)School of Mathematical and Physical Sciences, University of Sheffield, Sheffield S3 7RH, UK.
    
    \(^2\) Qlibri GmbH, 80337 Munich, Germany; Faculty of Physics, Ludwig-Maximilians-Universit{\"a}t M{\"u}nchen, 80799 Munich, Germany; Munich Center for Quantum Science and Technology, 80799 Munich, Germany.
    
  \(^*\)email: zstaylor1@sheffield.ac.uk\par
  \(^{**}\)email: a.tartakovskii@sheffield.ac.uk\par
  
    \vspace{1mm}
\endgroup

The exceptionally high in-plane refractive index, low sub-bandgap absorption, and strong optical anisotropy of WS\(_2\) make it a promising material platform for next-generation integrated circuits for nanophotonics. Its layered van der Waals structure further enables heterogeneous integration with silicon photonics and emerging two-dimensional optoelectronic materials. However, despite increasing interest in the waveguiding properties of WS\(_2\), experimental studies of wavelength-dependent modal confinement and attenuation remain limited. Additionally, though the extinction coefficient of WS\(_2\) is expected to be near-negligible beneath the bandgap, reported values span orders of magnitude, leading to large uncertainty in predicted modal decay lengths and wafer-scale integration feasibility. To resolve these ambiguities we perform hyperspectral cavity-enhanced imaging, determining high-resolution upper and lower bounds on the extinction coefficient of WS\(_2\) within the visible/NIR edge. We further employ scattering-type scanning near-field optical microscopy (s-SNOM) to probe TE\(_0\), TM\(_0\), and higher-order modes in both quasi-bulk and nanobeam WS\(_2\) waveguides across the 800–1400 nm spectral range, enabling identification of mode-specific trends in wavevector dispersion and loss. This work simultaneously assesses s-SNOM as a probe of waveguide performance, and we find that while absolute loss values depend on measurement geometry, s-SNOM reliably captures relative modal trends and provides upper bounds on propagation loss, supporting its use as a diagnostic tool for anisotropic waveguides. We further identify significant artefacts in nanobeam measurements arising from transverse interference and spatial sampling effects when the structure size approaches the excitation wavelength, which can shift extracted effective indices by up to \(\Delta n
_{eff}= -0.25\).

\section{Introduction}
\small
The pursuit of miniaturised, ultrafast and energy-efficient information systems has driven intense research into Photonic Integrated Circuits (PICs), exploiting light’s capacity to encode information via wavelength, phase, polarisation and amplitude to vastly improve data transfer rates whilst also maximising integration density \citep{1.2}. Compared to purely electronic systems, PICs offer high-speed, long-distance transmission immune to electromagnetic interference \citep{1.3}, enabling dense on-chip multiplexing \citep{1.3, 1.14} and providing the high bandwidth and efficiency essential in meeting the escalating demands of data-intensive fields such as artificial intelligence and machine learning \citep{1.6, 1.13}. For photonic materials, the key figure of merit is the refractive index \citep{1.8, 1.9}, which governs both the optical mode confinement (and thus component miniaturisation) and the degree of field overlap with the confining material, thereby influencing the strength of light-matter interactions crucial for sensing, switching and modulation applications. The imaginary component of this value (also known as the extinction coefficient, \(\kappa\)) dictates absorption, with near-negligible values targeted in waveguides to minimise loss. 

Thus far, the transparency of SiN at visible wavelengths \citep{SiN} and silicon in the near-infrared \citep{1.12}, combined with the latter’s high refractive index, has led these materials to dominate integrated photonics within their respective spectral windows \citep{1.14, 1.13}. These desirable features are further complemented by their compatibility with Complementary Metal-Oxide-Semiconductor (CMOS) \citep{1.15} processes which enable comparatively low-cost production of silicon-based waveguides with submicron cross-sections and single- to sub-nanometre sidewall roughness \citep{1.16, 1.17}. However, both material platforms encounter key limitations that restrict further scaling of integrated photonic systems. For example, the comparatively low refractive index of SiN requires larger waveguide dimensions to minimise optical losses, whereas for silicon the same transparency which facilitates low-loss light guiding in the NIR also prevents it from being able to detect light within that regime; as the indirect bandgap also prevents efficient light emission, silicon’s status as a poor emitter and detector necessitates the integration of alternative materials such as germanium \citep{1.18} or III-V semiconductors \citep{1.19} which can introduce lattice constant or thermal expansion coefficient mismatch \citep{1.20, 1.21} and therefore additional defects and strain, often with a deleterious impact upon device quality and performance. 

Van der Waals materials have been envisioned as a potential solution to both the integration challenges associated with silicon photonics and the refractive index limitations of SiN waveguides. The interlayer van der Waals forces promote strong adhesion to substrates and can mitigate the lattice mismatch requirement in silicon-based circuits \citep{1.23}. As potential active materials, few-layer transition metal dichalcogenides (TMDs) in particular exhibit a range of desirable features such as strongly bound excitons \citep{1.24, 1.25}, ultrahigh carrier mobility \citep{1.27}, thickness-tunable optical bandgaps \citep{1.28, 1.29} and high optical nonlinearity \citep{1.30}, features which have been utilised in a wide range of optoelectronic devices ranging from modulators \citep{1.32} and photodetectors \citep{1.33} to quantum LEDs \citep{1.34} and nanolasers \citep{1.35}. 

Beyond their role as active media, quasi-bulk TMDs show excellent promise as waveguides due to their layered crystal structure conferring very high in-plane refractive indices and low loss below the bandgap \citep{1.36, 1.37}. This strong geometrical anisotropy leads to highly anisotropic optical constants, with some crystals such as WS\(_2\) and MoS\(_2\) exhibiting a birefringence of \(\Delta n > 1\) in the vis-NIR regime \citep{1.38, 1.39}. WS\(_2\) is of particular interest because previous studies suggest that its exceptionally large birefringence can be achieved alongside low sub-bandgap absorption, implying high transparency throughout substantial portions of both the visible and near-infrared spectral ranges \citep{1.36, 1.37}, with this combination motivating the focus on WS\(_2\) throughout this work.

The heightened optical anisotropy of WS\(_2\) could be instrumental in reducing the scale of optical components to match with their electronic counterparts. In this material, deeply subwavelength light manipulation is enhanced near the bandgap, where strong excitonic resonances dominate the optical properties and highly confined exciton polariton (EP) modes are supported by extraordinarily thin crystals, allowing mode confinement in crystal dimensions down to approximately \(\lambda\)/40 \citep{1.40}, though this is at the expense of enhanced absorption near resonance. Recent studies have also observed vastly reduced crosstalk in TMD-based crystal waveguides comparative to silicon \citep{1.42, 1.43}, with a 20-fold decrease \citep{1.44} achievable in certain configurations that could facilitate improved component integration density.

Whilst these device-level advantages of WS\(_2\) are well established, the influence of strong optical anisotropy upon the extent of scattering loss within 2D material waveguides has not been made explicit. Additionally, reported WS\(_2\) sub-bandgap extinction coefficients demonstrate some variation across the literature \citep{1.37, 1.38, 1.43}, with small deviations in transparency leading to large changes in propagation loss on the \(\mu\)m scale. Most prior work has also focused only on isolated Transverse Electric (TE) modes, motivated by their high effective refractive index and reduced sensitivity to waveguide geometry within the form factors typical of conventional SOI photonics. Indeed, a key advantage of WS\(_2\) and other highly anisotropic TMDs is that the reduced refractive indices in the out-of-plane direction suppress the confinement of Transverse Magnetic (TM) modes, allowing support of a single TE mode \citep{1.43} over a wide range of nanobeam dimensions to further mitigate the effects of mode interference. However, several emerging photonic applications such as mode-division multiplexing \citep{1.46, 1.47}, hybrid photonic-magnetic systems \citep{1.48}, and platforms requiring strong out-of-plane dipole coupling \cite{1.49} demand careful consideration of both waveguide polarisations. Many of these applications rely on precise control of the wavelength-dependent modal dispersion, which governs phase matching, modal interference and the broadband spectral response of integrated photonic devices. For these use cases, it is therefore essential to understand how modal confinement and attenuation evolve with wavelength and geometry, as well as how interactions between supported modes develop across this parameter space.

In this work, we address these gaps by first providing high-resolution quantification of the extinction coefficient of WS\(_2\) in the sub-bandgap visible regime before evaluating scattering-type Scanning Near-Field Optical Microscopy (s-SNOM) as a probe of waveguiding quality. Unlike most prior studies, which typically examine guided modes at single wavelengths or treat modal behaviour in isolation, our approach provides systematic wavelength-dependent measurements across the 800–1400 nm range whilst resolving multiple modal behaviours (TE\(_0\), TM\(_0\) and TE\(_1\)).  

By applying square-root Lorentzian fitting to near-field fringe Fourier Transforms extracted from nanobeam and quasi-bulk WS\(_2\) waveguides, we extract modal dispersion and decay characteristics and identify how confinement-dependent effects govern the observed attenuation. While absolute propagation losses are influenced by s-SNOM-specific geometrical and coupling factors, the extracted trends provide insight into relative modal behaviour and the underlying mechanisms that control waveguiding performance in highly anisotropic crystals. These results demonstrate s-SNOM as a powerful experimental tool for benchmarking waveguiding quality in highly anisotropic van der Waals materials, capable of providing direct feedback towards waveguide design, fabrication optimisation and material selection in next-generation photonic integrated circuits. 

\section{Results and Discussion}

\begin{figure*}[h!]
	\centering
	\includegraphics[width=\linewidth]{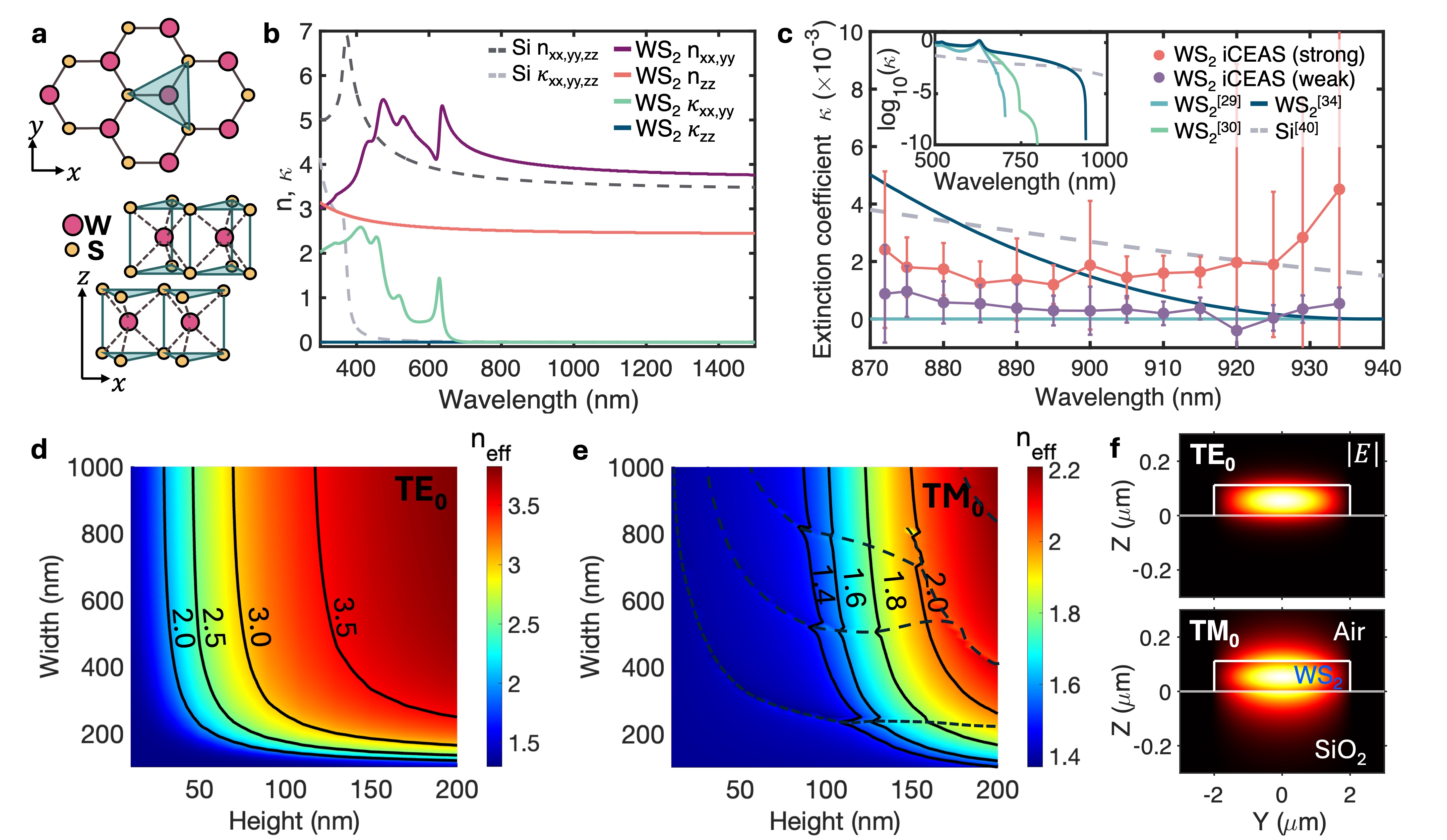}
	\small
	\caption{Optical waveguide quality of WS\(_2\) throughout the vis-NIR spectral window relative to silicon. a) atomic structure of WS\(_{2}\) as seen from above (top) and the side (bottom). b) Refractive indices in the vis-NIR window of WS\(_{2}\) and silicon sourced from \citep{1.38} and \citep{2.1}, respectively. c) Extinction  coefficient of a WS\(_2\) crystal measured via imaging Cavity-Enhanced Absorption Spectroscopy (iCEAS) within strongly and weakly absorbing local crystal regions compared with prior measurements of WS\(_2\) \citep{1.37, 1.38, 1.43} and Si \citep{2.1} \(\kappa\). A logarithmic plot of the extinction coefficient of WS\(_{2}\) and silicon is provided within the inset to emphasise the sharper drop in absorption estimated for WS\(_{2}\). d) and e) Simulated effective index maps of the (d) TE\(_{0}\) and (e) TM\(_{0}\) modes in WS\(_{2}\) on 1 µm SiO\(_{2}\) calculated in Lumerical FDE at \(\lambda = 800\) nm. Dotted lines are included within the TM\(_0\) mode plot to denote discontinuities due to coupling with higher-order TE modes. f) Example E-field profiles of the TE\(_{0}\) (top) and TM\(_{0}\) (bottom) modes in a 112 nm-thick, 4 \(\mu\)m width WS\(_2\) slab at \(\lambda = 800\) nm above a 1 \(\mu\)m SiO\(_2\) layer.}
	\label{fig:Fig1}
\end{figure*}

As shown in \Cref{fig:Fig1}a, WS\(_2\) consists of a stack of layered sheets in which tungsten atoms are covalently bonded into a hexagonal arrangement sandwiched between two trigonal layers of sulfur atoms, with each sheet held together by weak van der Waals forces. WS\(_2\) exhibits strong optical anisotropy arising from its layered geometry, with a narrow exciton resonance at 629 nm expressed in-plane in contrast with a flat, near-featureless refractive index out-of-plane (\Cref{fig:Fig1}b). Within the vis-NIR spectrum the disparity between \(n(o)\) and \(n(e)\) remains pronounced, though this is strongly wavelength-dependent; in the high-transparency region the birefringence shows significant dispersion from \(\Delta n = 1.61\) at 800 nm to \(\Delta n = 1.31\) at 1690 nm, driven primarily by the highly-dispersive in-plane refractive index. This \(n(o)\) dispersion represents an almost 2-fold increase over that observed in silicon \citep{1.38}.

The influence of this strong anisotropy on waveguiding quality can be understood via comparison with Lumerical FDE simulations of the effective indices of the confined TE\(_0\) and TM\(_0\) modes (\Cref{fig:Fig1}d-f) at excitation wavelength \(\lambda = 800\) nm. For both modes, \(n_{eff}\) increases asymptotically from the free-space index as the crystal width and thickness are increased, with the TE\(_0\) mode plateauing at \(n_{eff} \approx 3.9\) and the TM\(_0\) mode near \(n_{eff} \approx 2.3\). Importantly, the TM\(_0\) mode \(n_{eff}\) only surpasses the refractive index of the underlying SiO\(_2\) layer at WS\(_2\) crystal thicknesses above approximately 100 nm, whereas below this cutoff the field is unconfined and leaks evanescently into the substrate. Discontinuities in simulated \(n_{eff}\) due to strong hybridisation with higher-order TE modes are visible in the TM\(_0\) plot, marked by dashed lines and persisting throughout the guided and sub-cutoff (\(n_{eff}\) = 1.45) regions. These mode anti-crossing points in the dispersion are indicative of symmetry-allowed overlap with mode solutions with odd numbers of z and y nodes (TE\(_{10}\), TE\(_{30}\)).

In contrast, because the TE\(_0\) mode remains more strongly confined across a broader range of dimensions, phase matching with symmetric TM modes is altogether avoided and no markers of mode coupling are observed. Thus, different behaviours are anticipated from each mode profile: the TE\(_0\) mode, being solely dependent upon \(n(o)\), is expected to be highly dispersive, with its confinement quality varying significantly with wavelength. Meanwhile, the TM\(_0\) mode interacts primarily with the comparatively weakly dispersive \(n(e)\), resulting in reduced modal dispersion but enhanced evanescence and coupling sensitivity to wavelength-dependent confinement of higher-order TE modes. 

Beyond geometrical confinement, the waveguiding performance of these modes is also influenced by material absorption, which remains ambiguous in WS\(_2\) below the bandgap. \Cref{fig:Fig1}c compares reported extinction coefficients of WS\(_2\) collected from three separate sources \citep{1.37, 1.38, 1.43} against silicon \citep{2.1}, with a logarithmic scale inset applied for further clarity of absorption loss magnitude. Whilst silicon has undergone extensive, high-precision characterisation of its optical components through paired ellipsometry and spectrophotometry measurements \citep{2.1}, the three extinction coefficients of WS\(_2\) shown have been derived via ellipsometry and fitted with a Tauc-Lorentz oscillator model paired with UV, infinite frequency and Drude function terms. Although all three models predict that absorption in WS\(_2\) decreases more rapidly than in silicon above 800 nm (and is therefore often assumed to be negligible in this regime) the quantitative spread between them demonstrates the difficulty of reliably extracting minute attenuation via far-field methods. 

In order to obtain a quantitative value for the absorption in the actual geometries used under realistic conditions an imaging Cavity-Enhanced Absorption Spectroscopy (iCEAS) measurement was performed using a cavity microscope (``Qlibri Nano"). The high absorption sensitivity due to cavity enhancement enabled measurement of the absorption spectrum without reliance upon material modelling (\Cref{fig:Fig1}c, Supplementary Note \ref{supp:note1}). In addition to the variance in \(\kappa\) reported in literature, significant variability is observed even within a single sub-100 micron crystal produced via micro-mechanical exfoliation, which represents the fabrication method associated with the lowest defect density. The measured \(\kappa\) values exhibit an approximately linear trend and show good agreement with \citep{1.43} in the strong absorption iCEAS limit, and with \citep{1.37, 1.38} in the low absorption limit. These values correspond to an average absorption loss of 23 \(\pm\) 18 dB/mm in weakly absorbing crystal regions (Supplementary \Cref{fig:S1}) and up to 118 \(\pm\) 85 dB/mm in strongly absorbing areas, with the observed linear trend suggesting that comparable losses are likely to persist outside of the 872-934 nm regime throughout the near-infrared transparency window.

\begin{figure}[!t]
	\centering
	\small
	\includegraphics[width=\linewidth]{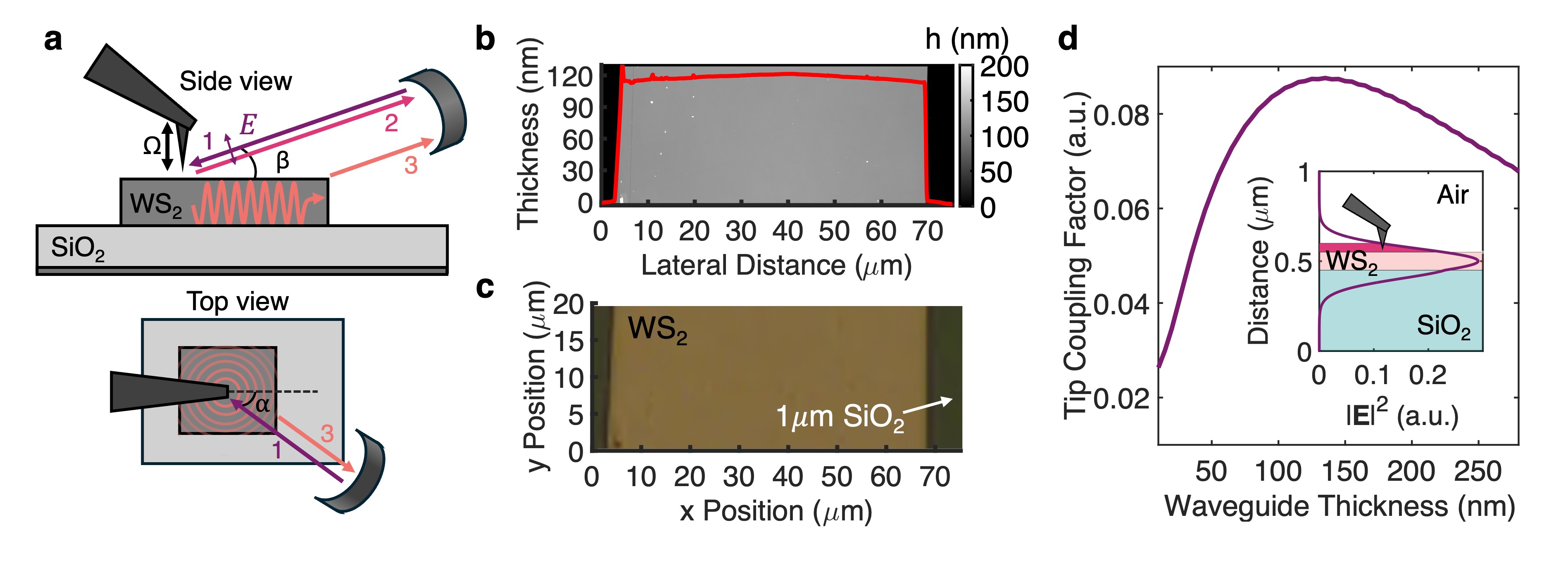}
	\caption{a) s-SNOM schematic showing side and top view. A metal-coated AFM tip is illuminated by a beam of laser light (path 1). This light is partially reflected directly to the detector from the tip (path 2), and partially coupled into the WS\(_{2}\) crystal via the plasmonic tip-sample interaction, propagating as a waveguided mode before scattering towards the detector at the crystal edge (path 3). Angles of the laser beam relative to the crystal edge, \(\alpha\) and \(\beta\), are indicated. b) Topography of quasi-bulk sample with a thickness of (112 ± 5) nm recorded via AFM demonstrating high uniformity and pristine edges. S-SNOM scans performed upon the RHS due to comparative lack of surface line and point defects. c) Bright-Field image of the quasi-bulk WS2 crystal on 1 µm SiO\(_{2}\). d) Tip coupling factor as a function of WS\(_{2}\) waveguide thickness, simulated in Lumerical and calculated as in \citep{2.4} assuming 50 nm tapping amplitude.}
	\label{fig:Fig2}
\end{figure}

The potential use of WS\(_2\) as a waveguide material in PICs requires probing of wavelength-dependent confinement quality and attenuation of both fundamental modes in WS\(_2\). To this end, an s-SNOM is employed to experimentally extract dispersion data. The working principle is demonstrated in \Cref{fig:Fig2}a: a non-polarised monochromatic laser beam is tightly focused onto the apex of a tapping metallic tip, inducing a localised plasmonic response that concentrates and enhances the optical field in the gap between tip and sample. The resulting near-field interaction generates two distinct scattering pathways. In the first, light is scattered directly from the tip-sample interaction, directly encoding the local near-field optical response of the crystal. For the second pathway, the light overcomes the momentum mismatch and is coupled into the crystal as a waveguided mode, which propagates as a cylindrical \citep{Mooshammer} wave towards the crystal edge and is subsequently scattered towards the parabolic mirror as a free-space photon. The interference between these two far-field signals at the detector produces a sinusoidally-varying fringe pattern in both the amplitude and phase images which encodes features of the propagating modes within the crystal. 

This edge-scattering of modes is expected to be the dominant fringe-producing interaction; edge reflection is strongly suppressed away from plasmon polariton resonances due to the proximity of the waveguide mode momenta to the free space wavevector, \(k_0\) \citep{2.2, 2.3}. Edge launching is also possible as the beam spot size is on the order of the free-space wavelength, but Fourier Transforms of the simultaneously-generated s-SNOM phase profiles show no peak signatures in the negative wavevector regime, implying that mode propagation is highly unidirectional and thus tip in-coupling has much higher efficiency. Where edge scattering is the dominant mechanism, the recorded \(m\)th-order s-SNOM amplitude signal (\(s_m\)) of these modes can be described by a sum of cylindrically-spreading decaying sinusoids, 

\begin{equation}
s_m(x) \propto \sum_{i=1}^{n} \frac{A_i}{\sqrt{x}} \, \sin\!\left[k_0 (n_{\mathrm{eff},i} - G)\, x \right]\, e^{-\gamma_i x},
\end{equation}

Where \(A_i\) defines the amplitude of the \(i\)th mode contribution, \(\gamma_i\) denotes its decay and the observed wavevector is given by the \(n_{eff, i} = k_{x,i}/k_0\) offset by a geometrical factor of \(G = cos(\beta)sin(\alpha)\) \citep{2.4}, where laser beam angles \(\alpha\) and \(\beta\) are described in \Cref{fig:Fig2}a. Upon Fourier Transform these fringes are converted into a series of square-root Lorentzian peaks that can collectively be described by the equation:

\begin{equation}
    Amp.[s_m] = Offset + \sum_{i=1}^{n}\frac{A_i}{\sqrt{1 + \left(\frac{\frac{k_x}{k_0}- n_{eff,i} + G}{\frac{\gamma_i}{2}}\right)^2}},
    \label{eq:eq2}
\end{equation}

where the \(Offset\) describes the noise floor of the Fourier Transform. By explicitly separating the signals via a Fourier Transform, and fitting the modes as individual peaks, the propagation decay, confinement and amplitude of each mode contribution in the fringe can be effectively isolated and compared.

To establish a baseline for confinement and loss measurements with the s-SNOM, it is important to begin with a system that minimises scattering sources of loss. In near-field optics, measured losses can be inflated by geometrical effects and the evanescent nature of the fields. Additionally, as light from waveguided modes can be scattered from any surface defect, a pristine, featureless crystal with sharp edges is required to minimise fringe contributions from surface roughness. Thus a large, quasi-bulk crystal (\(>\)100 \(\mu\)m in lateral dimensions) with minimal surface roughness (R\(_q\) = 2.2 nm, with correlation lengths \(L_{c,x} = 2.8\) \(\mu\)m and \(L_{c,y} = 1.4\)  \(\mu\)m, \Cref{fig:Fig2}b and c) was selected to approximate a homogeneous in-plane medium and to suppress backscattering contributions from rough sidewalls. Within this system the key remaining fringe loss mechanisms are due primarily to surface scattering, absorption and s-SNOM alignment-related losses.

Whilst initially it may seem prudent to select the thickest crystal possible to maximise confinement, the coupling strength of the TE\(_0\) mode into the crystal must be considered; the tip-induced dipole field is predominantly sensitive to the comparatively large vertical extent of the probe, and therefore couples most efficiently to the large out-of-plane evanescent tail of the TM\(_0\) mode. Though effective excitation of the TE\(_0\) mode is also achievable within the vis-NIR spectral range, the fringe amplitude of this contribution is more sensitive to confinement strength, with enhanced coupling occurring when the TE\(_0\) electric field overlap with the 50 nm tip oscillation region is maximised \citep{2.4}. In \Cref{fig:Fig2}d the simulated (Lumerical FDE) proportion of the field within the tip oscillation region is plotted as a function of waveguide thickness. This tip coupling factor reaches a maximum in the vicinity of 135 nm crystal thickness for excitation wavelength \(\lambda\) = 800 nm. Ultimately, the 112 ± 5 nm-thick crystal shown in \Cref{fig:Fig2}b provides a good compromise between fringe amplitude of the TE\(_0\) mode and sufficient confinement of the TM\(_0\) mode to distinguish it from the anticipated air mode peak, whilst negating strong coupling with the TE\(_1\) excitation throughout the vis-NIR regime.

As a final consideration, it is noted that the TM\(_0\) mode shows significant leakage into the SiO\(_2\) layer throughout the width-thickness parameter space shown in \Cref{fig:Fig1}e owing to the small index contrast. To suppress any additional leakage into the underlying silicon substrate, a 1 \(\mu\)m-thick SiO\(_2\) layer was employed for these measurements.

\begin{figure}[!t]
	\centering
	\small
	\includegraphics[width=\linewidth]{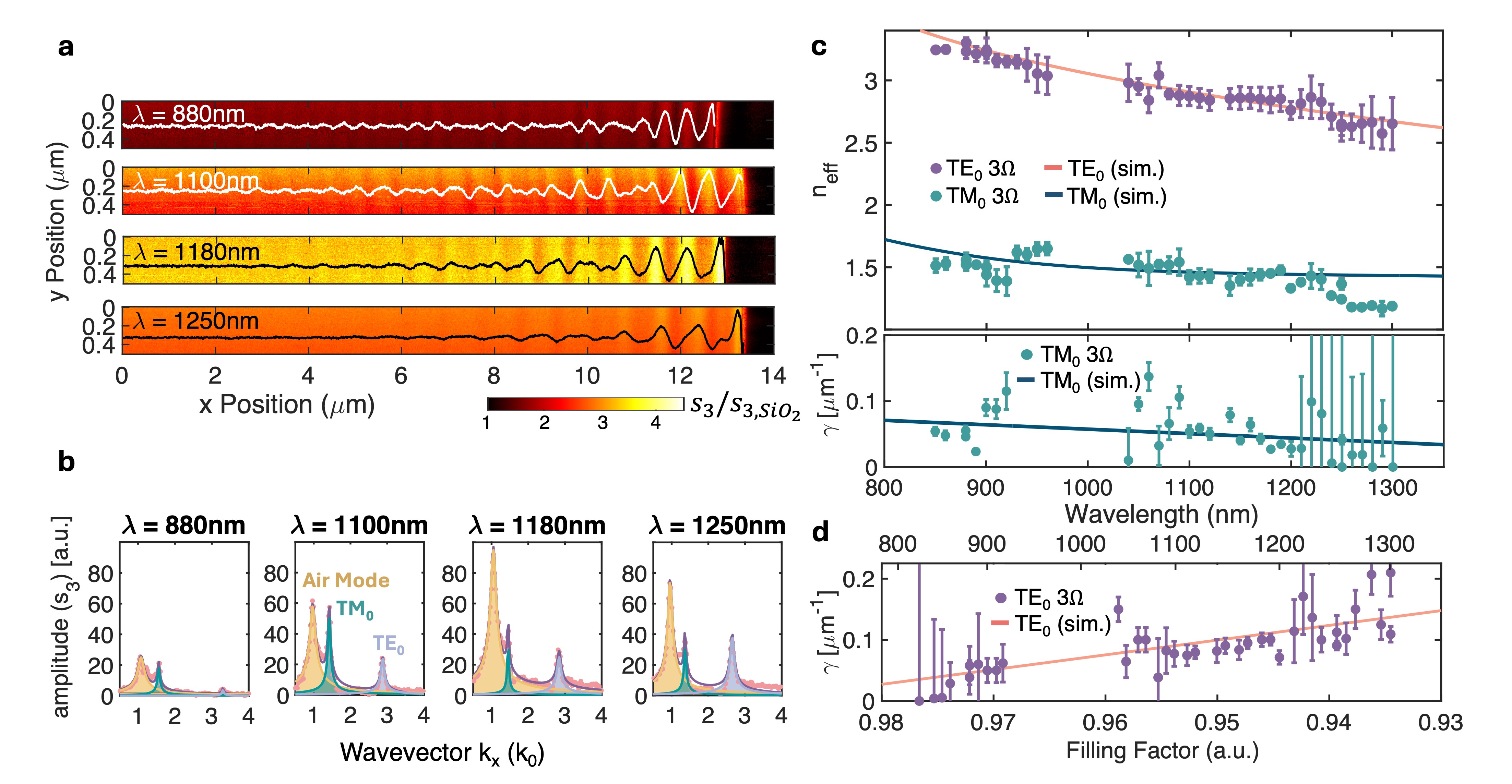}
	\caption{Nanoimaging of waveguide mode dispersion in WS\(_{2}\). a) SiO\(_{2}\)-normalised s-SNOM images of a quasi-bulk 112nm-thick WS\(_{2}\) crystal with fringe line profiles overlaid for a small selection of incident laser beam wavelengths (\(\lambda\)). b) Fitted Fourier Transforms of the line profiles in (a) demonstrating the presence of three key peaks; the air mode (yellow), a poorly-confined TM\(_{0}\) mode (cyan) and the TE\(_{0}\) mode (lilac). c) n\(_{eff}\) Fitting results for the TE\(_{0}\) and TM\(_{0}\) mode measured throughout the low-absorption vis-NIR window of WS\(_2\), with Lumerical MODE simulation results for a 4 \(\mu\)m-width, 112 nm-thick crystal included for comparison. The fitted decay constant of the TM\(_0\) mode is plotted in the lower panel d) Fitted TE\(_0\) decay constant plotted against filling factor, a measure of the proportion of the field distributed within the waveguide. 
}
	\label{fig:Fig3}
\end{figure}

\Cref{fig:Fig3}a presents representative s-SNOM amplitude images of the quasi-bulk crystal shown in \Cref{fig:Fig2}b-c at selected wavelengths. All images were demodulated at the third harmonic of the tapping tip, providing a good compromise between background suppression and signal-to-noise ratio, and normalised to the SiO\(_2\) substrate, whose negligible dispersion within the allocated regime (\(\Delta n <\) 0.009 \citep{2.5}) enables direct comparison between fringe profiles. Immediately it can be seen from the fringe profiles overlaid in \Cref{fig:Fig3}a that the propagation length of the included modes decreases rapidly upon increasing wavelength; whilst fringes are clearly visible almost 10 \(\mu\)m from the crystal edge at 880 nm excitation, they vanish within 7 \(\mu\)m at 1250 nm. This occurs because as the excitation wavelength increases the TM\(_0\) mode reaches cutoff and increasingly leaks into the substrate, thus the relatively lower-amplitude and increasingly poorly-confined TE\(_0\) mode becomes the dominant long-range contribution.

Example Fourier Transforms are shown in \Cref{fig:Fig3}b, where three contributing peaks can be identified. As expected from prior crystal characterisation, two waveguide modes are present (TE\(_0\) and TM\(_0\)) alongside the unconfined air mode, which typically occupies wavevector values of \(k_x/k_0 = 1\) after the geometrical offset is applied. Whilst the wavevector position of this contribution is usually stable, it represents unconfined light and as such a larger fraction of the field resides within the tip’s sweep area, resulting in stronger coupling and predominance in the fringe waveform. However, the broad linewidths (\(\gamma_{avg} = 0.15\)) of this air mode visible in \Cref{fig:Fig3}b indicate that, owing to its unconfined nature, it decays much more rapidly from the crystal edge than the TE\(_0\) and TM\(_0\) modes also present in the Fourier spectra. As higher-order modes begin to approach confinement they are often expressed as shoulders upon this peak and shift its position slightly as shown in Supplementary \Cref{fig:S3}. 

The fringe Fourier spectra across all wavelengths were fitted with (\ref{eq:eq2}) to extract \(n_{eff}\), \(\gamma\) and \(A\) for the fundamental guided modes (\Cref{fig:Fig3}c,d; Supplementary \Cref{fig:S3}-6). The experimentally-derived effective indices demonstrate good agreement with the simulated slab waveguide index data. As expected, \(n_{eff}\) decreases with wavelength in parallel with the refractive index dispersion, with this effect being more pronounced in the TE\(_0\) mode (\(\Delta n_{eff, TE0} = 0.98\)) than the TM\(_0\) mode (\(\Delta n_{eff, TM0} = 0.29\)) throughout the 800-1300 nm range, reflecting the highly dispersive nature of the refractive index along the ordinary plane. Above approximately 1200 nm excitation wavelength the effective index matching of the TM\(_0\) mode between simulation and experiment is poor, and this is primarily due both to a reduction in peak amplitude and to partial melding of the TM\(_0\) peak with the air mode. These trends indicate practical design considerations: for TE\(_0\), the sharp decrease in effective index with wavelength implies reduced broadband operation and greater broadband sensitivity to fabrication errors, while for TM\(_0\), larger waveguide dimensions are required to maintain confinement, increasing susceptibility to dispersion from higher-order TE modes.

Interestingly, the modal decay constants exhibit opposing wavelength-dependent trends. The TE\(_0\) mode shows increasing decay with wavelength, while the TM\(_0\) mode exhibits a weak decrease. This behaviour arises from their distinct confinement characteristics; the TE\(_0\) mode remains guided but becomes progressively less confined, increasing its field overlap with surface defects at the air-WS\(_2\) and WS\(_2\)-SiO\(_2\) interfaces. Conversely, the TM\(_0\) mode quickly approaches cutoff, such that most of its energy is distributed throughout the underlying SiO\(_2\) and Si layers. This reduced air-WS\(_2\) field intensity results in lower scattering losses and hence slightly longer propagation due to low sensitivity to the WS\(_2\) waveguide structure.

For the TE\(_0\) mode, this relationship between confinement and decay is further supported by comparison with the simulated filling factor \(\Gamma\), defined as:

\begin{equation}
\Gamma
=
\frac{
\int_{\text{core}}
\varepsilon(x,y)\, |E(x,y)|^{2} \, \mathrm{d}x\,\mathrm{d}y
}{
\int_{\text{total}}
\varepsilon(x,y)\, |E(x,y)|^{2} \, \mathrm{d}x\,\mathrm{d}y
},
\end{equation}

where \(\varepsilon(x,y)\) denotes the in-plane permittivity component and \(E(x,y)\) the in-plane field. This \(\Gamma\) variable quantifies the fraction of a mode's energy confined within the waveguide core. A strong negative linear correlation between \(\Gamma\) and the extracted decay constant is observed (\Cref{fig:Fig3}d; Supplementary \Cref{fig:S5}), confirming that increased field leakage directly contributes to enhanced propagation loss. This trend is consistent across demodulation orders. No comparable correlation is observed for the TM\(_0\) mode due to its substantially different field distribution, though the weak trend in decay is also consistently reproduced across demodulation orders (Supplementary \Cref{fig:S5}).

Lineaer fits to the wavelength-dependent propagation losses, obtained by converting the fitted decay constants to dB/mm, resulted in a gradient of -0.29 dB/mm/nm for the TM\(_0\) mode and 0.94 dB/mm/nm for the TE\(_0\) mode at 3rd order demodulation. In absolute terms, the extracted decay lengths correspond to propagation losses between 150-1600 dB/mm for both modes. In comparison with the prior measurements of absorption via iCEAS, dB/mm loss in the 880-935 nm range is 240-270 for the TM\(_0\) mode and 200-260 for the TE\(_0\) mode according to the fitting. Thus, the combined effects of s-SNOM geometry, enhanced field leakage, and surface scattering represent an apparent elevation of 137 \(\pm\) 90 dB/mm for the TM\(_0\) mode and 112 \(\pm\) 86 dB/mm for the TE\(_0\) mode where the strongly-absorbing scenario is considered. However, these values significantly overestimate the intrinsic propagation loss.

It is important to note at this stage that whilst s-SNOM measurement conditions and alignment were maintained as fixed as possible, these measurements took place over multiple sessions. The observation of clear, reproducible decay trends that mirror simulated relationships suggests that the geometrical collection efficiency remained relatively stable despite variations in wavelength and tip condition. Furthermore, the consistent scaling of the extracted decay gradients by a common, enhanced factor indicates that geometrical contributions act primarily as a systematic scaling of the measured signal, consistent with a multiplicative effect on the observed decay rather than a purely additive contribution. 

If this s-SNOM multiplicative factor indeed remains stable throughout the measured range, and the extinction measured via iCEAS is accurate within the visible window, then a multiplicative factor accounting for s-SNOM geometrical considerations may be approximated via comparison between measured and expected loss in the visible window as (2.16\(\pm\)1.57)dB/mm for the TM\(_0\) mode and (1.95\(\pm\)1.4)dB/mm for the TE\(_0\) mode (again assuming that the strongly-absorbing iCEAS scenario is the case), such that true optical losses may instead follow a -0.13\(\pm\) 0.09 dB/mm/nm (TM\(_0\)) and 0.48\(\pm\)0.35 dB/mm/nm (TE\(_0\)) relationship. Thus, whilst absolute loss values obtained via s-SNOM may not be directly representative of intrinsic material attenuation, they can still reveal how field overlap with interfaces and scattering sites governs the relative loss trends.

Having examined mode character in a highly homogeneous system, the next step was to assess multimode propagation and loss dynamics within a series of thinner waveguides representative of structures employed in PICs. In such structures, field overlap with the boundaries of the waveguides is increased and surface defects become strong scatterers of propagating light, enhancing backscattering and intermodal coupling effects. Note that because the system is more sensitive to field distribution in both transverse directions, modes are named according to the number of field maxima along transverse axis, such that TE\(_{00}\) denotes the fundamental TE mode.

\begin{figure}[!t]
	\centering
	\small
	\includegraphics[width=\linewidth]{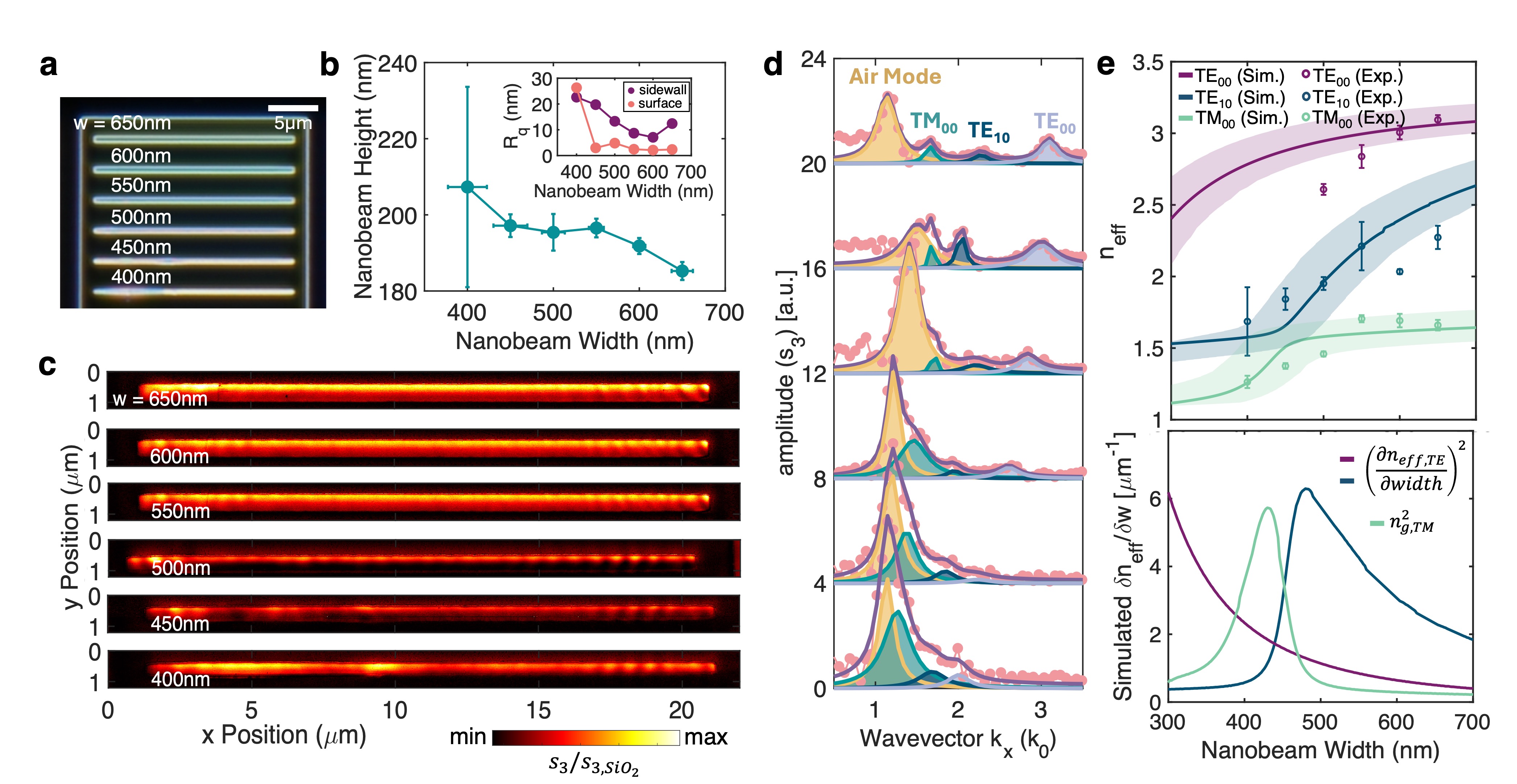}
	\caption{Characterisation and s-SNOM fringe measurements of 200 nm-thick WS\(_{2}\) nanobeams. a) Bright-Field (BF) images of WS\(_{2}\) nanobeams fabricated via EBL and RIE. b) Nanobeam height and width measured from AFM images, with RMS roughness (R\(_{q}\)) applied as error bars. Measured surface and sidewall roughness is provided in an inset, exhibiting the high surface quality of exfoliated TMD crystals. c) Third-order amplitude images of the nanobeams normalised to the SiO\(_{2}\) substrate at \(\lambda\) = 1300 nm. d) Fourier Transforms of the RHS fringes from (c). e) (top) \(n_{eff}\) obtained from Fourier Transform fits compared with simulated width dispersions. Upper and lower bounds in this value are represented by the shaded regions calculated using nominal widths incorporating maximum observed surface and sidewall dimension uncertainty. (bottom) width-dependent backscattering coefficient calculated as in \citep{2.6}. Simulation data in (e) obtained in Lumerical FDE.}
	\label{fig:Fig4}
\end{figure}

For this purpose an array of nanobeams was fabricated via Electron Beam Lithography (EBL) and Reactive Ion Etching (RIE) on a WS\(_2\) flake which had been micro-mechanically exfoliated onto a 290 nm-thick SiO\(_2\)/Si substrate. The nanobeams were designed to access the regime of strong mixing between the TE\(_{10}\) and TM\(_{00}\) modes whilst maintaining near-unity transmission of a strongly confined TE\(_{00}\) mode under ideal (smooth sidewall) conditions. To achieve this, the waveguide thickness was fixed at 200 nm, providing strong confinement of the TE\(_{00}\) and TE\(_{10}\) modes and minimising leakage of the TM\(_{00}\) mode into the underlying silicon substrate. Nanobeam length was further fixed at 20 \(\mu\)m to minimise interference between mode fringes launched from opposing waveguide edges, whilst a 2 \(\mu\)m spacing between adjacent structures suppressed optical crosstalk. With these parameters chosen, nanobeam width was varied from 650 nm to 400 nm in 50 nm increments, enabling tuning of the modal dispersion through the regime of strong TE\(_{10}\)-TM\(_{00}\) mixing at excitation wavelengths of 1200 and 1300 nm. The resulting width-variable nanobeams are visible in the dark-field images of \Cref{fig:Fig4}a and the AFM scans in Supplementary \Cref{fig:S8}. 

The fabrication process inevitably introduces sidewall roughness in TMD nanobeams, producing structural perturbations that break the ideal symmetry of the waveguide and act as scattering centres for propagating modes. Surface roughness remains low due to the inherent stability and cleavage properties of WS\(_2\) as shown in \Cref{fig:Fig4}b, where surface RMS roughness generally takes values between 2-4 nm for waveguides without significant defects. Sidewall roughness, meanwhile, is larger by an order of magnitude owing primarily to RIE.

\Cref{fig:Fig4}c shows the 3rd-order demodulated s-SNOM scans at an excitation wavelength of \(\lambda\) = 1300 nm of the nanobeam samples exhibited in \Cref{fig:Fig4}a. Fringes can be identified on both ends of the waveguides, though each decays rapidly and typically extends 10 \(\mu\)m or less into the crystal, such that line profiles can be taken and Fourier Transformed from the right edge of the waveguide with minimal contribution from the fringe on the opposite end.

Enhanced field leakage from unconfined modes is also visible in the form of increased amplitude signal outside of the waveguide boundaries, a feature that becomes increasingly pronounced with reducing nanobeam width. These patterns are expected to originate primarily from leakage of the TM\(_{00}\) mode into the air surrounding the nanobeams; the boundary conditions of TM-polarised light cause a sharp increase of field intensity at the in-plane WS\(_{2}\)-air boundary which decays exponentially with distance from the waveguide edge as visible in \Cref{fig:Fig4}c. The decay from the waveguide edge, \(\gamma_{ev}\), of this tail is given by \(\gamma_{ev} = k_0\sqrt{n^2_{eff} - n^2_{air}}\), such that the decay reduces concomitantly with mode wavevector and hence width, explaining the increasing strength of leakage observed from thinner nanobeams. Fringe effects are also visible within these evanescent tails, an artefact that originates from continued interference with tip-scattered light at the detector, with fringe modulation outside the waveguide exclusively replicating the air mode frequency rather than the beat-like effects observed within the waveguide.

This increasing lateral mode leakage is also described within the frequency analysis of the fringe waveform. The Fourier Transforms taken from the full waveguide y-position average of the right edge waveform are visible in Fig. 4d. Therein, four peaks are visible, and upon FDE mode analysis can be matched to the air mode, fundamental TM\(_{00}\) and TE\(_{00}\) modes, and TE\(_{10}\) mode. The air mode shows increasing dominance of the FT spectrum upon reducing width whereas TM\(_{00}\) mode quickly transitions from existing entirely separately into appearing consistently as a shoulder of this air mode peak as the field increasingly leaks into the underlying SiO\(_2\) and Si substrate layers. Due to the melding of both peaks individual mode decays and amplitudes are difficult to disentangle and quantify, but tend to occupy values on the order of 0.2-0.3 \(\mu\)m\(^{-1}\), corresponding to observed losses of 870-1300 dB/mm. Thus, the TM and air mode fringe decay rapidly from the nanobeam edge, leading to dominance of the TE\(_{00}\) and TE\(_{10}\) mode fringes within the inner extent of the waveguide.

The simulated dispersion of the waveguide modes as a function of reducing nominal nanobeam width is plotted in \Cref{fig:Fig4}e. The TE\(_{00}\) peak is of particular interest, showing significant deviation from simulation (\(\Delta n_{eff} = n_{eff, exp} - n_{eff,sim} = -0.25\)) with nanobeam width. This drop in \(n_{eff}\) takes on an almost linear character in contrast with the asymptotic curve expected from the simulated data. This trend is replicated at excitation wavelength \(\lambda\) = 1200 nm (Supplementary \Cref{fig:S9}) and cannot be solely attributed to wavevector reduction from backscattering, as the error area shown in the upper panel of \Cref{fig:Fig4}e already accounts for nanobeam width and height variations arising from surface and sidewall roughness. Furthermore, the extent of backscattering within the measured nanobeam width range is expected to be much higher for the TE\(_{10}\) mode as exhibited in the lower panel of \Cref{fig:Fig4}e, yet no similar trend is observed for that mode. 

\begin{figure}[!t]
	\centering
	\small
	\includegraphics[width=\linewidth]{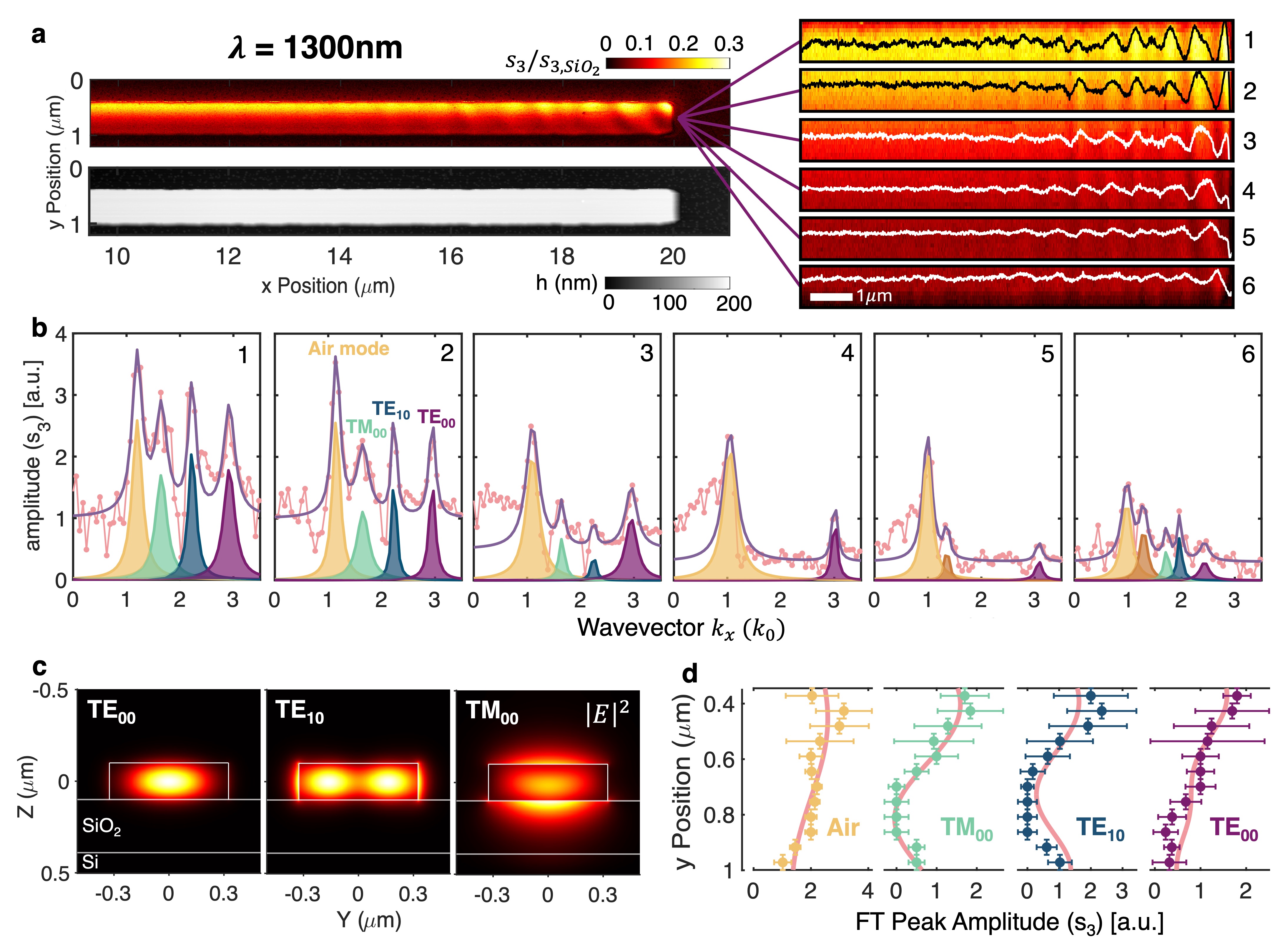}
	\caption{Waveguided mode analysis within a single nanobeam. a) Normalised 3rd order amplitude and AFM measurement of the 650nm-width nanobeam at \(\lambda = 1300\)nm. \(s_3\) Fringe measurements are decomposed according to y position on the right with profiles overlaid, demonstrating changing fringe pattern properties dependent upon Y position. b) Fitted Fourier transforms of decomposed nanobeam fringes 1-6 from panel a), showing the presence of an air mode (yellow) and TM\(_00\) (cyan), TE\(_{00}\) (purple), TE\(_{10}\) (blue) modes. An additional orange peak is included to account for poorly-confined substrate modes. c) Mode profiles of the confined TE\(_{00}\), TE\(_{10}\) and TM\(_{00}\) modes. d) FT peak amplitudes as a function of y position.}
	\label{fig:Fig5}
\end{figure}

An alternative explanation may lie in the technique of waveform extraction. Within this paper, fringes are extracted by Fourier Transforming the average over the full width of the nanobeam. However, decomposing the fringe waveform by y-position (\Cref{fig:Fig5}a) reveals that fringe characteristics vary significantly along the short axis of the nanobeam. Specifically, the s\(_3\) FT amplitude signals of the TE\(_{10}\) and TM\(_{00}\) modes are strongly suppressed near the bottom of the nanobeam (\Cref{fig:Fig5}b). This effect arises from fringe formation along the short axis and is consistent with the amplitude variations visible in \Cref{fig:Fig4}c. Interference between modes scattered from the nanobeam’s short edges produces a sinusoidal modulation of each mode’s amplitude along the y-axis, meaning that any given fringe extracted from a single position may underrepresent or entirely suppress certain modes.

At any x-y position from the edges of a waveguide of width \(w\), the amplitude signal is given by:

\begin{equation}
s_m(x) \propto \sum_{i=1}^{n} \left(\frac{A_{i,1}}{\sqrt{x}} \, \sin\!\left[k_0 (n_{\mathrm{eff},i} - G_x)\, x \right]\, e^{-\gamma_i x} + \frac{A_{i,2}}{\sqrt{y}}sin\left[\frac{\pi y(n_{eff,i} + G_y)}{w}\right] + \frac{A_{i,3}}{\sqrt{y}}sin\left[\frac{\pi y(G_y-n_{eff,i})}{w}\right]\right),
\end{equation}

Where \(G_x\) and \(G_y\) represent the geometrical offset factor from the long and short axes, respectively. 
These transverse patterns are visualised explicitly in \Cref{fig:Fig5}d, where the amplitude of the FT peak of each confined mode across the full nanobeam transverse axis shows excellent agreement with (4). It is important to note that these patterns do not necessarily represent quasi-Fabry-Perot modes as the wavevectors remain close to the free-space wavevector and thus reflectivity is low. Indeed, reflectivity would be expected to be 0 for the air mode yet the pattern is still observed in that case. 

As each mode exhibits a finite linewidth in Fourier space, variation in y-position effectively applies a spatial weighting function that selectively samples different regions of each modal peak. Consequently, the extracted peak position is not fixed but can shift depending on the spatial profile of the measurement. As a result, in this nanobeam example the TE\(_{00}\) mode position shifts slightly by \(\Delta n_{eff} = 0.2\). This shifting may account for the deviation of experimental results from simulation in \Cref{fig:Fig4}e and can also be recognised in other highly lossy signals such as the air mode. Taking the average over the waveguide does not guarantee removal of this artefact as the transverse sinusoidal fringes do not necessarily complete a full cycle for each mode, and taking the line profile from the centre as in \citep{1.43} may actively suppress expression of some modes in waveguides where the extent of the short axis is comparable to the excitation wavelength. This also further complicates loss measurements from the FWHM of each FT peak. Accurate extraction of mode wavevector, loss and amplitude should therefore require decomposition and fitting of the modes along the short axis.

\section{Conclusions}
In summary, this work has demonstrated that s-SNOM enables systematic, wavelength-resolved characterisation of guided modes in highly anisotropic WS\(_2\) waveguides. First, we characterised the previously ambiguous extinction coefficient of WS\(_2\) in the 872-934 nm spectral window, providing upper and lower bounds for the intrinsic absorption properties of waveguided modes in this range and determining that the loss is of the same order of magnitude as silicon, confirming that whilst WS\(_2\) remains an adequate candidate for micron-scale visible light waveguiding, it is unsuitable for wafer-scale waveguiding within that wavelength range. By combining near-field imaging with Fourier analysis, we extract modal dispersion and relative decay trends, revealing the direct relationship between confinement and attenuation for the TE\(_0\) mode and distinct behaviour for the TM\(_0\) mode near cutoff. The observed trends provide insight into the scattering mechanisms governing waveguide performance and indicate that s-SNOM-generated geometrical contributions to the measured decay act as a stable scaling factor under consistent experimental conditions. Whilst these effects are empirical and specific to the measurement configuration, their stability suggests that, with appropriate calibration against reference structures with known loss, s-SNOM could provide a route to estimating propagation decay in comparable systems without the need for dedicated coupling structures, alongside the pre-established ability of this method to derive modal wavevector. Extension to nanobeam structures further reveals the role of transverse interference and spatial sampling in distorting extracted modal properties, emphasising the need for position-resolved analysis in sub-wavelength-scale structures.

\newpage

\setcounter{secnumdepth}{0}

\subsection{Methods}

\subsubsection{Sample Preparation}

\textbf{WS\(_2\) Exfoliation} WS\(_2\) flakes were mechanically exfoliated from a bulk crystal (HQ-Graphene) upon a hot plate heated to 60\(^{\circ}\)C to facilitate stronger adhesion to the \(1\mu\)m (quasi-bulk sample) and 290 nm (nanobeam sample) SiO\(_2\) on Si substrates. Large (\(>60\) \(\mu\)m width) flakes were identified through optical microscopy and their thicknesses and surface roughness measured with a Dimension Icon-PT Closed Loop Scanning Probe Microscope operating in tapping mode. The locations and crystal axes of approximately 200 nm-thick WS\(_2\) samples upon the 290 nm-thick SiO\(_2\) on Si substrate were noted in preparation for nanobeam patterning.

\textbf{Electron Beam Lithography and Reactive Ion Etching.} To produce the nanobeams the samples were spin-coated with a 400 nm-thick layer of positive resist (ARP13 AllResist GmbH) at 4000 rpm for 30 seconds, followed by a 2-minute bake at 180\(^{\circ}\)C. Electron beam lithography was performed using a Raith GmbH Voyager system, operating at an accelerating voltage of 50 kV and a beam current of 1.026 nA, to define arrays of parallel nanobeams with varying widths and a 1 \(\mu\)m clearance around each structure in the resist. The patterned samples were then developed by immersion in xylene and IPA, selectively removing the exposed resist to form a mask for reactive ion etching. This etching step was conducted for 40 seconds at 0.14 mbar pressure and a DC bias of 135 V using a fluorinated gas mixture of CHF\(_3\) and SF\(_6\). Any remaining resist was then removed by submerging the sample in warm 1165 resist remover for 20 minutes followed by further cleaning in acetone and IPA for 10 and 5 minutes, respectively.

\subsubsection{Characterisation}

\noindent \textbf{iCEAS} The iCEAS method is based on an optical microcavity formed by a planar mirror and a concave micro mirror at the end of an optical fibre. Light introduced into the cavity via the optical fibre undergoes multiple reflections between the highly reflective mirrors, which strongly enhances its interaction with any nanoscale sample placed on the planar mirror. This amplification allows the detection of extremely weak optical absorption that would otherwise be difficult to measure. The microcavity design also confines the optical mode to a small lateral region on the sample, enabling diffraction-limited 3.7 \(\mu\)m spatial resolution. By laterally moving the two mirrors relative to each other, the sample can be scanned, allowing the generation of spatial maps of optical extinction that encompass both absorption and scattering. Repeating this process over the wavelength range of 872-934 nm produces hyperspectral maps, providing a detailed characterisation of the material’s optical response at the nanoscale. 

To extract extinction from the absorption values measured from a 100 nm-thick WS\(_2\) crystal with this technique, the measured transmission was fitted against simulations of the cavity that applied the real component of the WS\(_2\) refractive indices reported by \citep{1.43}. Additionally, due to residual contamination from the micro-mechanical exfoliation technique, a global loss term of 150 ppm was incorporated.

\noindent \color{black}\textbf{s-SNOM} s-SNOM measurements were performed under ambient conditions using a commercial neaSCOPE from Attocube Systems AG. The setup comprises a tapping-mode AFM illuminated by monochromatic light sourced from a 140 fs pulsed Ti:Sapphire laser with an output range between 340 to 1600 nm (Chameleon Compact OPO-Vis by Coherent). This light is split into two branches by a Michelson interferometer, where one branch contains an oscillating reference mirror and the other focuses the light by a parabolic mirror onto the metal-coated tip (ARROW-EFM from Apex Probes, which possesses a PtIr coating, ~25 nm tip radius and 75 kHz driving frequency), whereupon a plasmonic response is induced resulting in strong field confinement and enhancement between tip and sample. Light backscattered from this interaction, alongside light scattering from the crystal edge, is collected by the same parabolic mirror and recombined with light from the interferometer reference arm at the detector. By superimposing the tip-sample signal with the sinusoidal reference field from the oscillating mirror, the field amplitude and phase can be isolated. By further demodulating the signal amplitude s\(_3\) at the third harmonic of the tapping frequency, a good balance between signal-to-noise ratio and background suppression is achieved in the vis-NIR regime.

For all discrete wavelength measurements the tip-tapping amplitude was restricted to 45-55 nm to minimise fringe amplitude dependency. Crystals were consistently oriented such that the launching edge (or the short edge in nanobeam measurements) was at an in-plane angle of \( \alpha = 45^{\circ}\) and out-of-plane angle of \( \beta = 40^{\circ}\) with respect to the incident laser beam. 

\noindent \textbf{FDTD and FDE Simulations} Finite-Difference Time-Domain and Finite-Difference Eigenmode simulations were performed using Lumerical software. Ellipsometry data from sources \citep{2.1} and \citep{1.43} provided the refractive indices used to model mode propagation in homogeneous Si and WS\(_2\) waveguides upon SiO\(_2\) substrates. A port source was used to launch and solve mode eigenvalue equations to obtain the relevant field profiles and effective refractive indices, whereas a mode source was paired with an index and DFT monitor to calculate filling factor.

\subsubsection{Contributions}
ZST performed s-SNOM measurements with contributions from AJK and TCP. ZST analysed the data with contributions from LMH, OW and AIT. SS and FF carried out iCEAS measurements and analysed the data. ZST and XH made samples. ZST wrote the manuscript with contributions from all co-authors. AIT and LRW supervised the project.

%\subsubsection{Funding}
%This work was supported by the Engineering and Physical Sciences Research Council (EPSRC).

\subsubsection{Acknowledgements}
ZST, AIT, XH, TCP, and AJK thank EPSRC grants EP/V007696/1 and EP/V006975/1. LMH, OTW, AIT and LRW thank EPSRC grants EP/V026496/1 and EP/S030751/1.

%For the purpose of open access, the author has applied a Creative Commons Attribution (CC BY) licence to any Author Accepted Manuscript version arising.
%
%\subsubsection{Declaration}
%All authors declare no financial or non-financial competing interests. 
%
%\subsubsection{Data Availability Statement}
%The data that support the findings of this study are available from the corresponding author upon reasonable request. 

\newpage
\bibliographystyle{unsrt}
\bibliography{thesis.bib}

\newpage
\setcounter{figure}{0}
\renewcommand{\thefigure}{S\arabic{figure}}

\subsection{Supplementary Notes}
\subsubsection{iCEAS}
\label{supp:note1}

Imaging Cavity-Enhanced Absorption Spectroscopy (iCEAS) is applied to produce hyperspectral extinction maps of quasi-bulk WS\(_2\) flakes. For this method, sample flakes are exfoliated upon a specialised macroscopic distributed Bragg reflector (DBR) mirror, which forms the planar base of the cavity. The iCEAS measurement is performed using a scanning cavity microscope, formed when a concave DBR micromirror fabricated at the end facet of an optical fibre is positioned above the planar mirror. The vertical separation between the mirrors and lateral position of the fibre can be adjusted via piezoelectric actuators, with the conferred fine spatial control allowing for detailed scanning and support of multi-wavelength light sources including the 872-934 nm range applied in this paper. 

When the cavity is illuminated through the optical fibre, the large finesse (\(F \approx 20,000\)) achieved through the high reflectivity of the DBR mirrors excites Fabry-Perot modes at discrete cavity lengths which satisfy the cavity resonance condition. Light undergoes multiple reflections between the mirrors, interacting with the sample multiple times before being transmitted through the sample mirror to the detector. This repeated interaction amplifies otherwise weak absorption within the sample, enhancing features in extinction which would typically be unresolvable with a single transmission pass. Due again to the high finesse, these Fabry-Perot resonances have very narrow linewidths and hence further enhanced sensitivity to sample absorption.

Absorption (\(A\)) encompasses scattering from both sample boundaries and surface roughness alongside the intrinsic optical absorption within the sample, and is extracted from relative transmission through the planar mirror. At resonance, total transmission (\(T_{cavity+sample}\)) is given:

\setcounter{equation}{0}
\renewcommand{\theequation}{S\arabic{equation}}

\begin{equation}
T_{cavity+sample} = \frac{4T_{m1}T_{m2+s}}{(T_{m1} + T_{m2+s} + L_{m1} + L_{m2}+2A)^2},
\label{eq: eq_1}
\end{equation}

where the total transmission (\(T\)) and loss (\(L\)) of the concave mirror (\(m1\)) and the planar mirror with the sample on top (\(m2+s\)) are given by \(T_{m1}\), \(L_{m1}\), \(T_{m2+s}\) and \(L_{m2+s}\), respectively. This can then be normalised against resonant transmission in an empty cavity, 

\begin{equation}
T_{cavity} = \frac{4T_{m1}T_{m2}}{(T_{m1} + T_{m2} + L_{m1} + L_{m2})^2},
\label{eq: eq_1}
\end{equation}

where all \(m2\) terms now represent a location upon the planar mirror absent of the sample. Relative transmission, \(T_{r}\), is thus given:

\begin{equation}
T_{r} = \frac{T_{cavity+sample}}{T_{cavity}} = \frac{T_{m2+s}}{T_{m2}}\frac{(T_{m1} + T_{m2} + L_{m1} + L_{m2})^2}{(T_{m1} + T_{m2+s} + L_{m1} + L_{m2}+2A)^2}.
\label{eq: eq_1}
\end{equation}

 This can then be rearranged to extract the absorption of the sample from a single pass, \(A_{sp}\):

\begin{equation}
A_{sp}=\frac{1}{2f.e.}\left(\sqrt{\frac{T_{m2+s}}{T_{r}T_{m2}}}(T_{m1} + T_{m2} + L_{m1} + L_{m2})-(T_{m1} + T_{m2+s} + L_{m1} + L_{m2})\right).
\label{eq: eq_1}
\end{equation}

Note that an additional field effect term (\(f.e.\)) is included to account for the wavelength-dependent field amplitude of the standing wave within the cavity. Finally, the total round-trip transmission and loss terms are expressed in terms of cavity finesse using \(F =2\pi/(T+L)\),

\begin{equation}
A_{sp}=\frac{1}{f.e.}\left(\sqrt{\frac{T_{m2+s}}{{T_{r}}T_{m2}}}\frac{\pi}{F_{cavity}} - \frac{\pi}{F_{cavity+sample}}\right).
\label{eq: eq_1}
\end{equation}

The finesse can be both measured experimentally from resonance linewidths and simulated from mirror reflectivity; in this paper, simulated finesse values based on the real part of the sample refractive index sourced from \citep{1.43} are used to isolate crystal absorption. An additional global loss of 150 ppm is applied within calculation of these finesse values to account for scattering due to residue left behind during the TMD exfoliation process. With \(A_{sp}\) thus extracted, this can be related to extinction \(\kappa\) via the Beer-Lambert law, wherein intensity loss due to crystal absorption is given by:

\begin{equation}
\frac{I(z)}{I_0} = e^{-\frac{4\pi\kappa z}{\lambda_0}}.
\label{eq: eq_1}
\end{equation}

By setting \(z\) as the thickness of the crystal sample (100 nm in this case), the intensity terms can be normalised as \(I_0 = 1\) and \(I(z)=1-A_{sp}\). Subsequently inverting this relationship allows extinction to be extracted as:

\begin{equation}
\kappa = \frac{\lambda_0}{4\pi z}\ln\left(\frac{1}{1-A_{sp}}\right).
\label{eq: eq_1}
\end{equation}

The concave fibre mirror confines the cavity mode to a Gaussian spot with a 3.7 \(\mu\)m radius in the visible spectral range, such that when the fibre mirror is rastered across the planar mirror surface a diffraction-limited \(\mu\)m-resolution spatial map of material extinction can be produced. This can be repeated for all \(\lambda_0\) for hyperspectral mapping, allowing extraction of dispersive absorption effects.

\begin{figure}[!t]
	\centering
	\small
	\includegraphics[width=\linewidth]{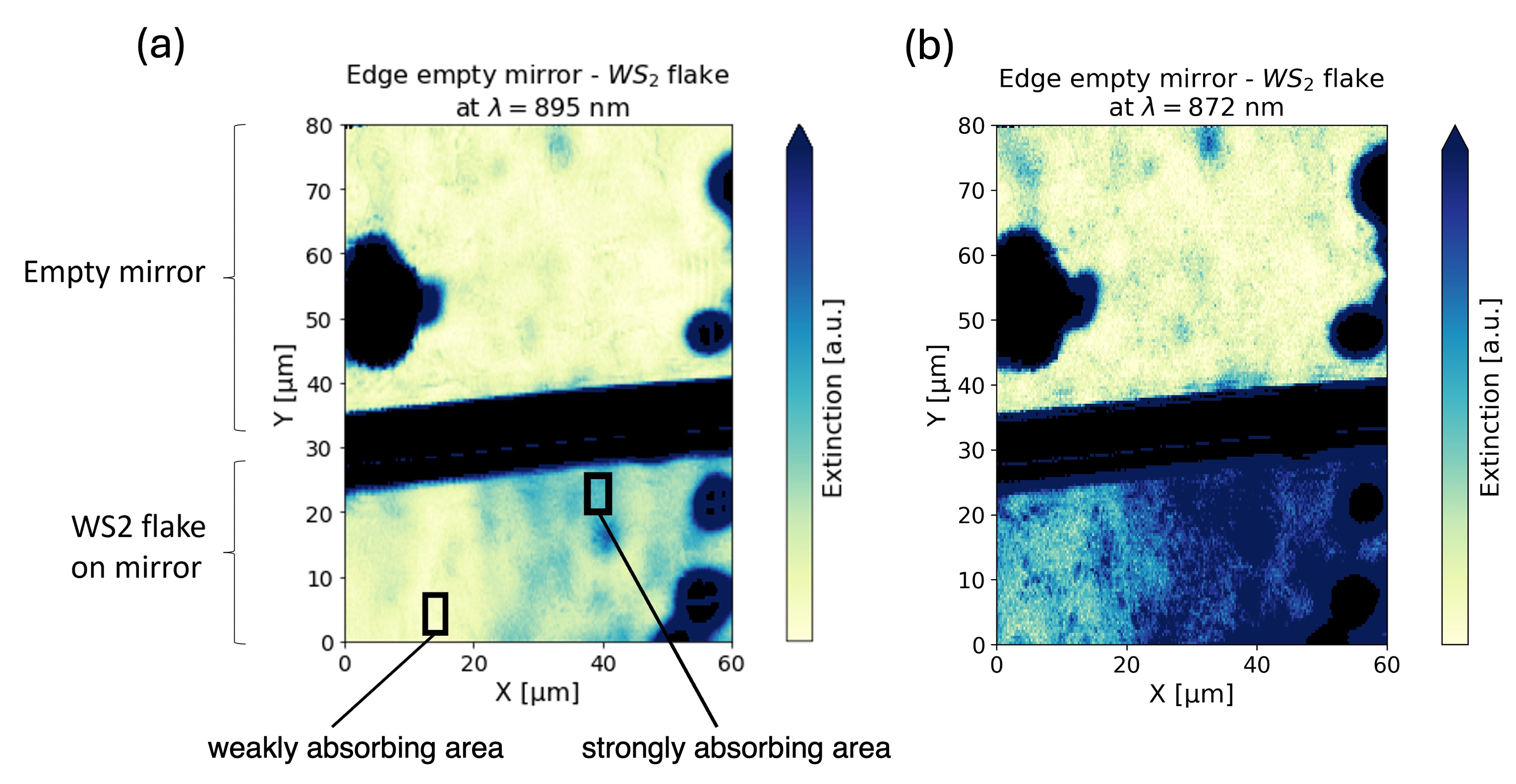}
	\caption{iCEAS extinction maps of a WS\(_2\) flake upon a mirror at a) 895 nm and b) 872 nm, from which the absorption values in Fig. 1c were calculated. 
}
	\label{fig:S1}
\end{figure}

\begin{figure}[!t]
	\centering
	\small
	\includegraphics[width=\linewidth]{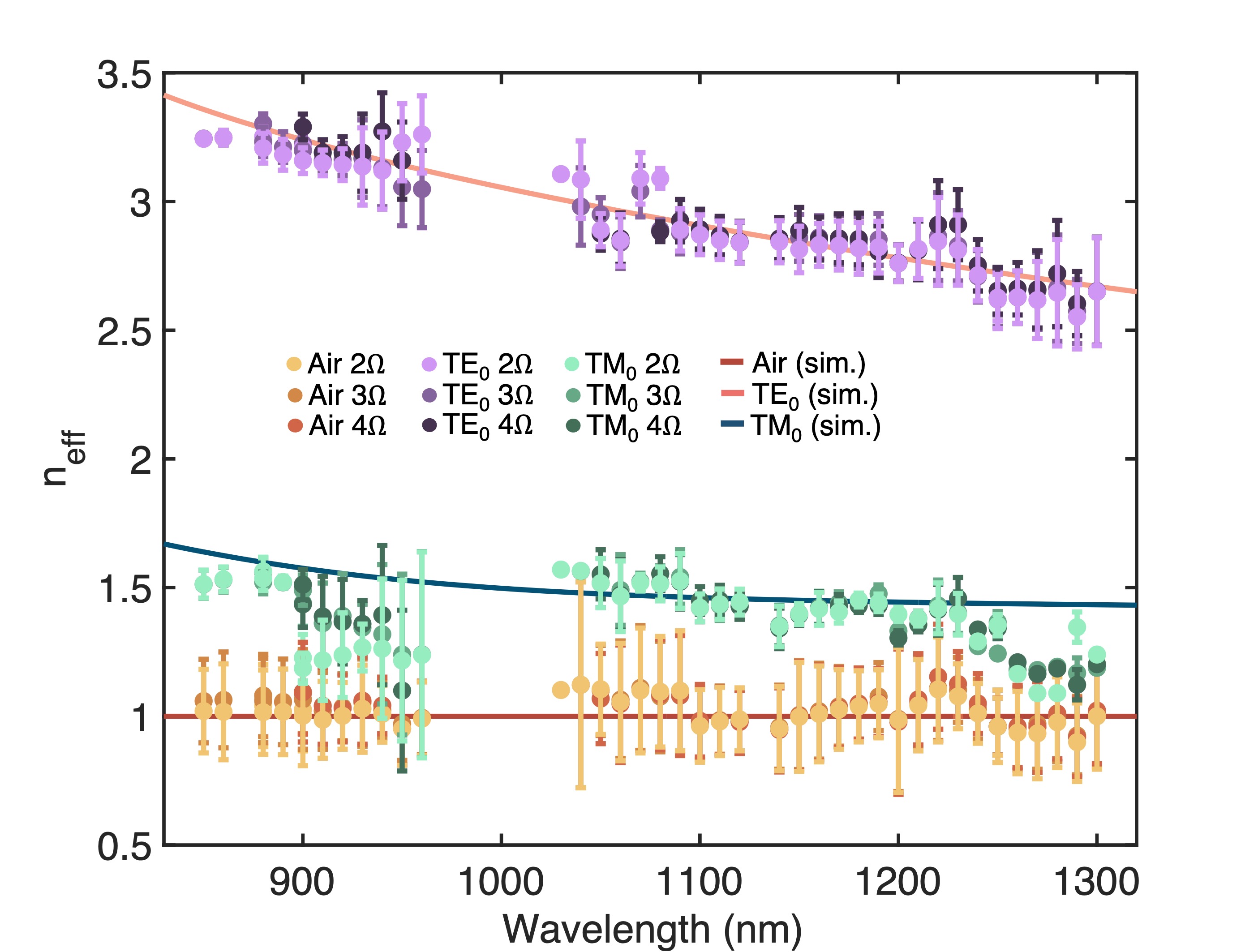}
	\caption{Normalised and corrected waveguide mode dispersion in a 112 nm-thick WS\(_2\) flake including fitting of the air mode. Experimental results are fitted against Lumerical FDE simulations of the waveguide effective index.
}
	\label{fig:S3}
\end{figure}

\begin{figure}[!t]
	\centering
	\small
	\includegraphics[width=\linewidth]{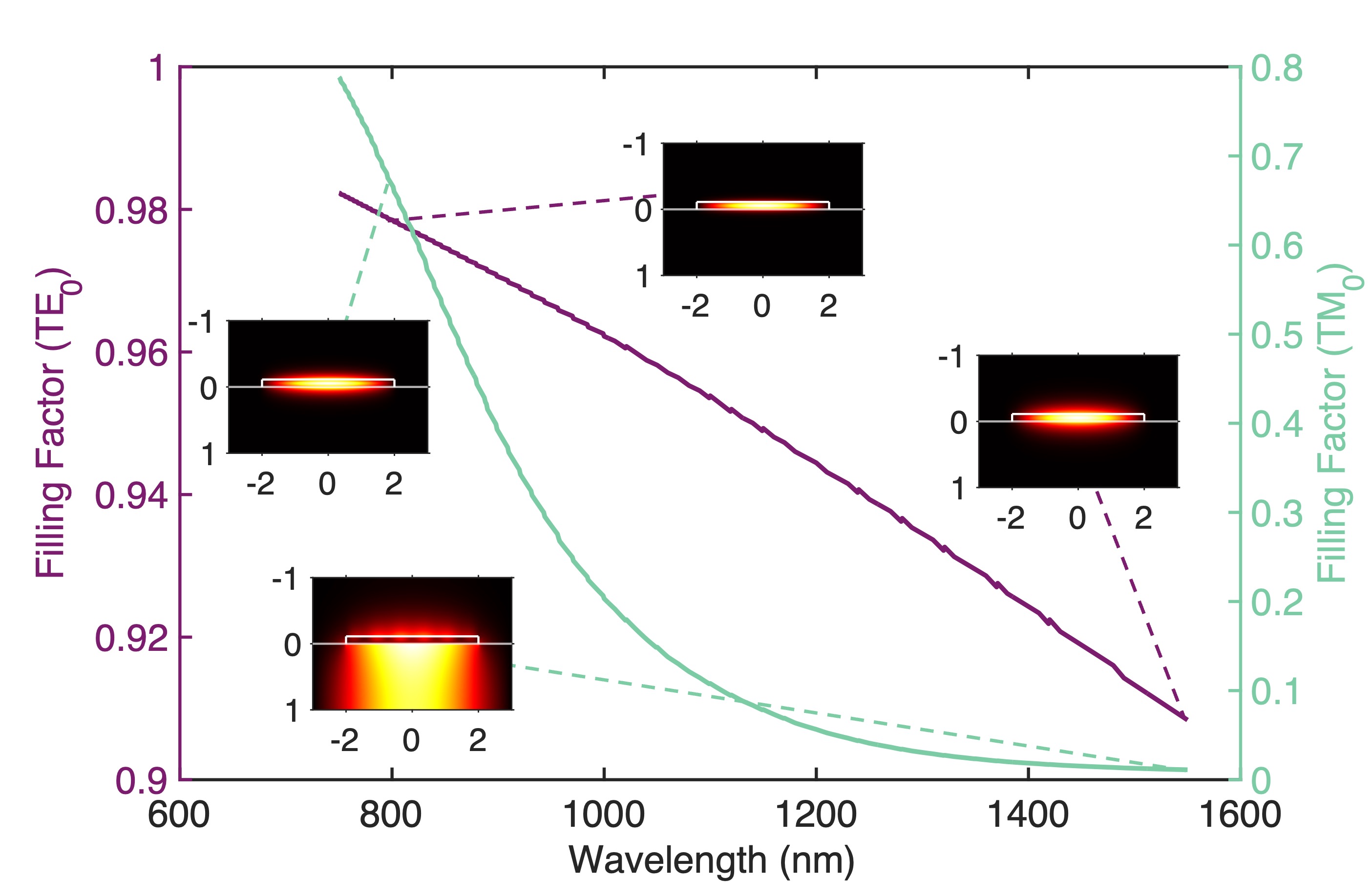}
	\caption{Simulated filling factor of the TE\(_0\) and TM\(_0\) mode (Eq. 3) calculated for a 112 nm-thick crystal in Lumerical FDTD. Field profiles of each mode calculated via FDTD at 800 nm and 1550 nm are provided as insets. 
}
	\label{fig:S4}
\end{figure}

\begin{figure}[!t]
	\centering
	\small
	\includegraphics[width=\linewidth]{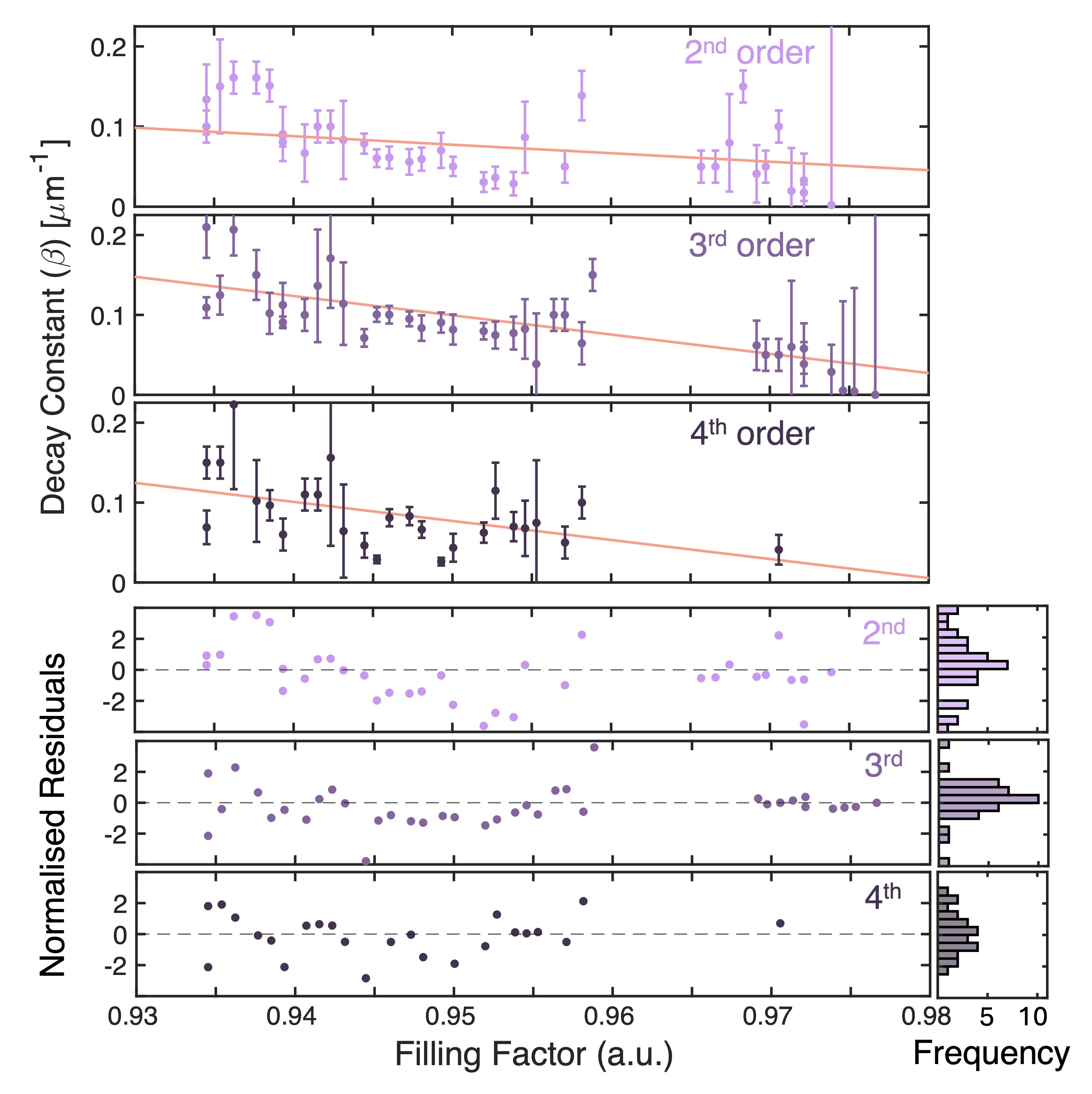}
	\caption{Decay constant plotted against TE\(_0\) mode filling factor showing a negative linear relationship across all low-background demodulation orders recorded. Corresponding normalised residuals included in lower panels alongside histograms demonstrating approximately Gaussian distribution.}
	\label{fig:S5}
\end{figure}

\begin{figure}[!t]
	\centering
	\small
	\includegraphics[width=\linewidth]{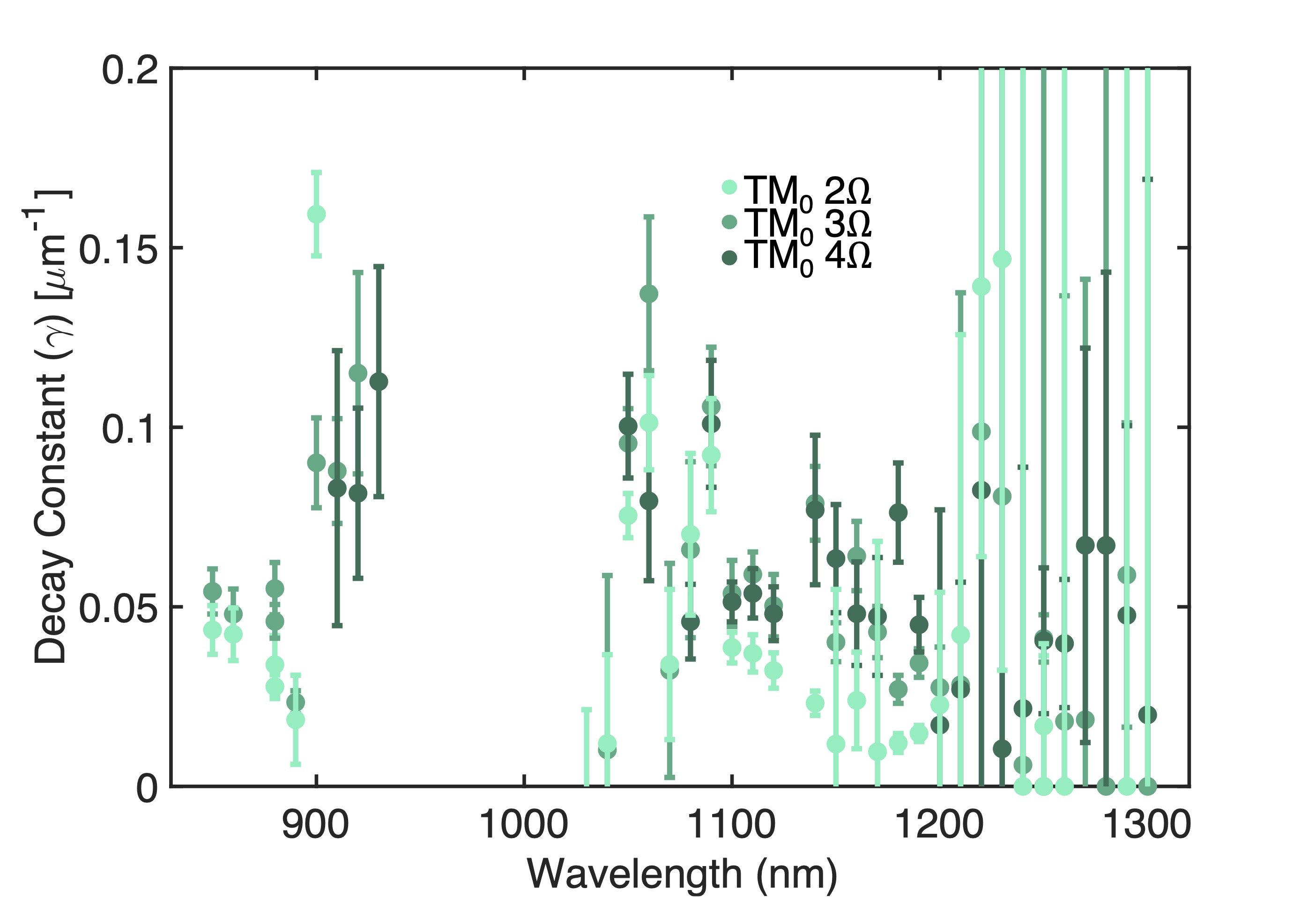}
	\caption{TM\(_0\) mode decay coefficients extracted from FT peak fitting shown for all demodulations, exhibiting a negative wavelength trend.
}
	\label{fig:S6}
\end{figure}

\begin{figure}[!t]
	\centering
	\small
	\includegraphics[width=\linewidth]{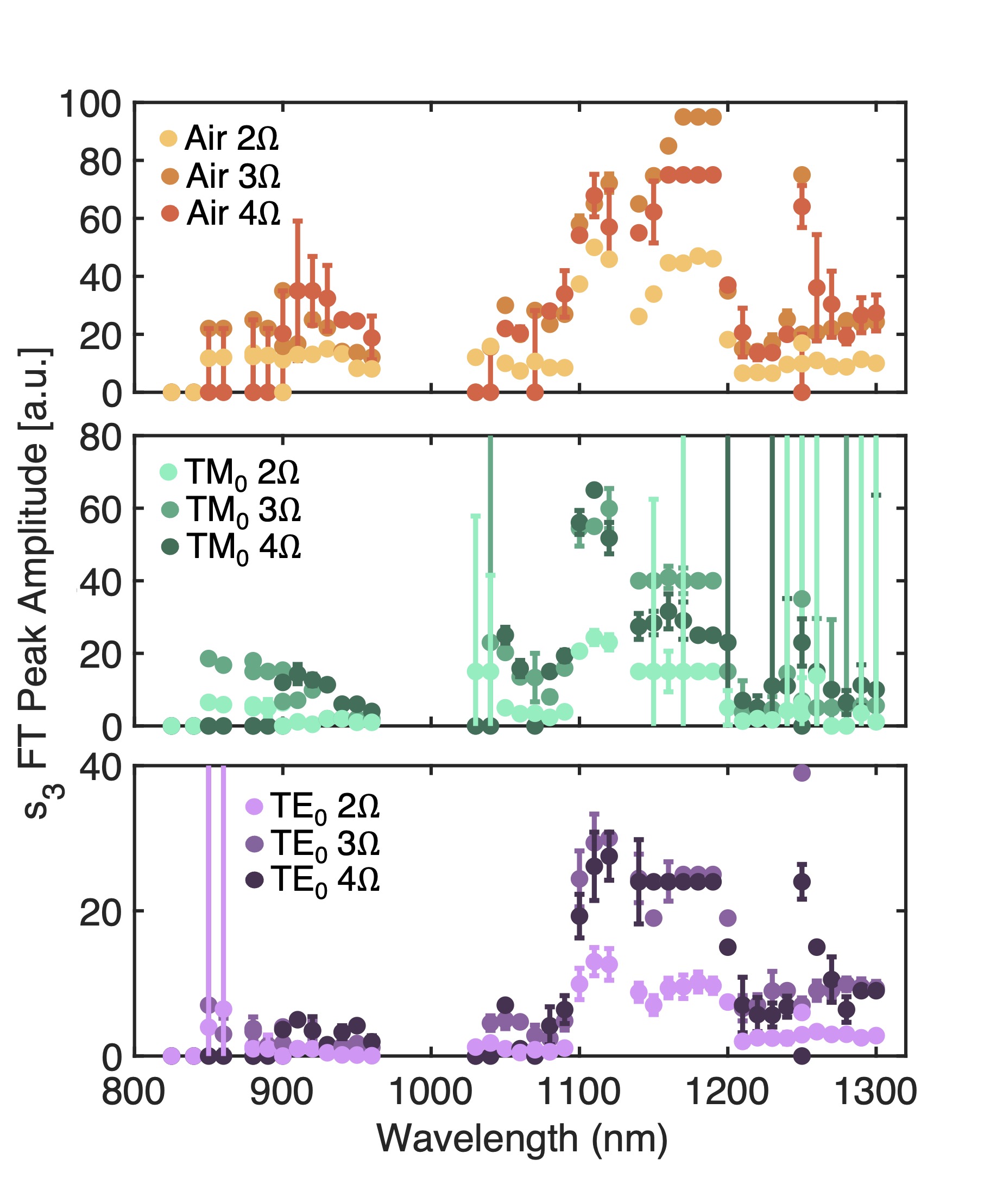}
	\caption{Fitted Fourier Transform peak amplitudes of the air, TM\(_0\) and TE\(_0\) modes extracted from fringe waveforms. All three modes exhibit broadly similar spectral trends, with the expected amplitude contrast; the air mode dominates the response across the measured range whereas the TM\(_0\) modes consistently show stronger coupling than the TE\(_0\) modes. Despite normalisation of the original waveforms to the SiO\(_2\) substrate, the amplitudes express comparable spectral structure irrespective of mode character and do not scale proportionally with expected field leakage into the tip-sample interaction region, suggesting an additional dependence within the fringe generation process. Error bars are included only for cases in which the amplitude parameter is well-constrained.
}
	\label{fig:S7}
\end{figure}

\begin{figure}[!t]
	\centering
	\small
	\includegraphics[width=\linewidth]{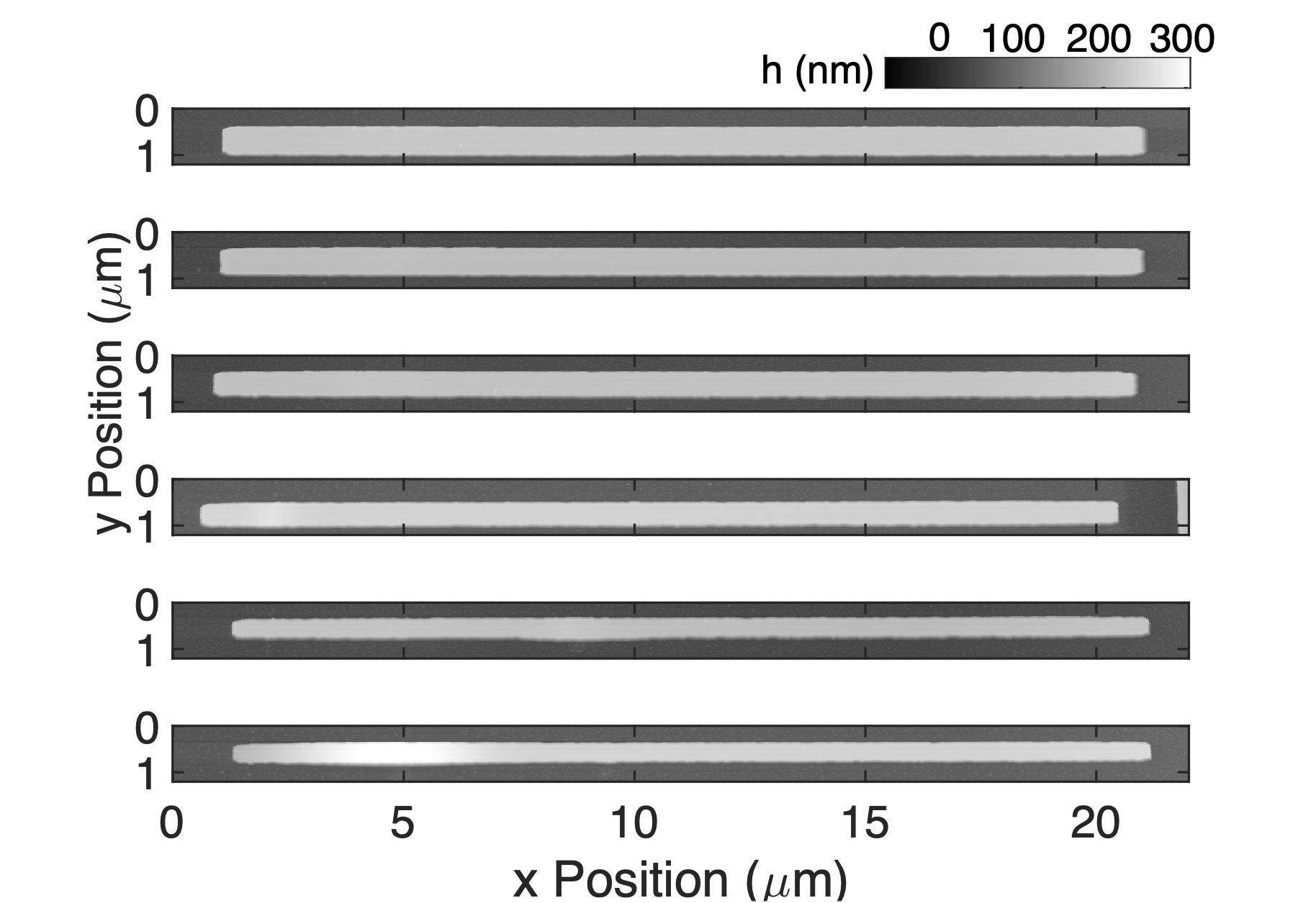}
	\caption{AFM images of EBL-fabricated nanobeams demonstrating width progression within sample sets. Each nanobeam is 20 \(\mu\)m long, with widths reducing from 650 nm (top panel) in 50 nm increments to 400 nm (bottom panel). The average nanobeam thickness is 195 nm as indicated by the included colorbar.}
	\label{fig:S8}
\end{figure}

\begin{figure}[!t]
	\centering
	\small
	\includegraphics[width=0.5\linewidth]{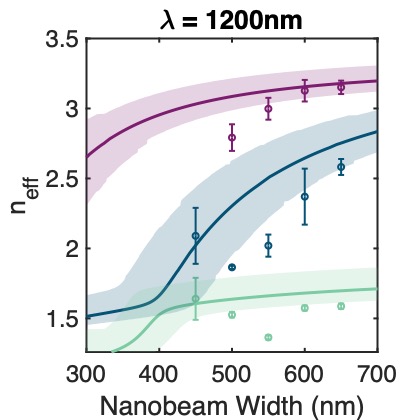}
	\caption{ Effective index (\(n_{eff}\)) at \(\lambda\) = 800 nm extracted from Fourier transform of fringe waveforms measured from full-width averaging over the waveguide, compared with simulated width-dependent dispersion. The shaded regions indicate upper and lower bounds calculated from nominal waveguide widths including the maximum observed surface and sidewall dimensional uncertainties. A similar trend in deviation from the asymptotic width dispersion as that seen in Fig. 4 is observed at this wavelength. Purple, TE\(_{00}\); blue, TE\(_{10}\); green, TM\(_{00}\).}
	\label{fig:S9}
\end{figure}

\begin{figure}[!t]
	\centering
	\small
	\includegraphics[width=\linewidth]{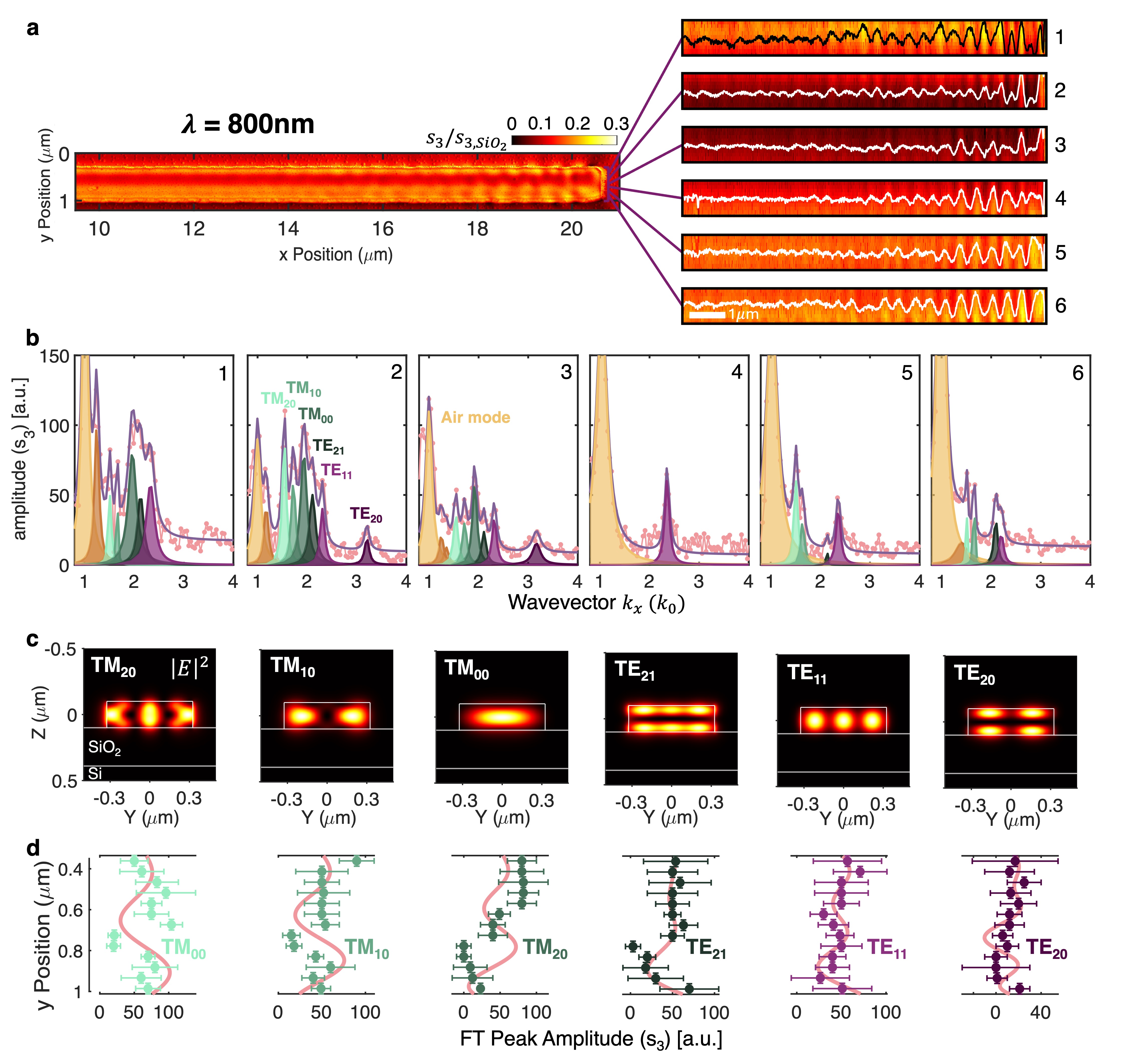}
	\caption{Waveguided mode analysis within single nanobeams repeated at \(\lambda = 1200\)nm. a) Normalised 3rd order amplitude and AFM measurement of a 650 nm-width nanobeam, the same as analysed in Fig. 5, with fringe measurements decomposed according to y position on the right. b) Fitted Fourier Transforms of decomposed nanobeam fringes. c) Mode profiles of the confined modes identified via peak fitting. The waveguide near-field map was divided into 12 profiles to allow for higher resolution mapping of the amplitudes of confined modes. d) FT peak amplitudes as a function of y position fitted according to Eq. 4.}
	\label{fig:S10}
\end{figure}

\end{document}